\documentclass[12pt,a4paper]{article}

\usepackage[utf8]{inputenc}
\usepackage[english]{babel}
\usepackage{cite}
\usepackage{xspace}
\usepackage{bbm}

\usepackage{amsmath,amssymb,amsfonts} 

\usepackage{tensind}
\tensordelimiter{`}

\def\*#1 {\mathbf{#1}}

\usepackage{fullpage}
\numberwithin{equation}{section}
\usepackage[bottom]{footmisc} 
\usepackage[margin=\parindent,font=small,labelfont=md,
labelsep=endash]{caption}

\makeatletter
\g@addto@macro\bfseries{\boldmath}
\makeatother

\usepackage[hidelinks]{hyperref}

\newenvironment{formula}%
{\begin{equation}\begin{aligned}\relax}%
{\end{aligned}\end{equation}\ignorespacesafterend}

\renewcommand{\H}{\ensuremath{\mathrm{H}}\xspace}
\newcommand{\G}{\ensuremath{\mathrm{G}}\xspace}

\makeatletter
\def\_#1 {{\underline{#1}}}
\makeatother

\def\oe{\mbox{\ensuremath{\mathring{e}}}}
\def\ow{\mbox{\ensuremath{\mathring{w}}}}

\newcommand{\cE}{{\mathcal{E}}\xspace}

\newcommand{\cG}{{\mathcal{G}}\xspace}
\newcommand{\cH}{{\mathcal{H}}\xspace}

\newcommand{\cM}{{\mathcal{M}}\xspace}

\newcommand{\cP}{{\mathcal{P}}\xspace}

\newcommand{\bbL}{{\mathbb{L}}\xspace}

\newcommand{\bbP}{{\mathbb{P}}\xspace}

\newcommand{\bbR}{{\mathbb{R}}\xspace}

\newcommand{\bbT}{{\mathbb{T}}\xspace}

\newcommand{\bbZ}{{\mathbb{Z}}\xspace}

\newcommand{\bbone}{\mathbbm{1}\xspace} 

\def\Gg{\ensuremath{\mathrm{G}}\xspace}
\def\Hg{\ensuremath{\mathrm{H}}\xspace}

\newcommand{\Eseven}{\ensuremath{\mathrm{E}_{7(7)}}\xspace}
\newcommand{\Esix}{\ensuremath{\mathrm{E}_{6(6)}}\xspace}
\newcommand{\eseven}{\ensuremath{\mathfrak{e}_{7(7)}}\xspace}

\def\SL(#1){\ensuremath{\mathrm{SL}(#1)}}
\def\PSL(#1){\ensuremath{\mathrm{PSL}(#1)}}
\def\GL(#1){\ensuremath{\mathrm{GL}(#1)}}
\def\Sp(#1){\ensuremath{\mathrm{Sp}(#1)}}
\def\USp(#1){\ensuremath{\mathrm{USp}(#1)}}
\def\SU(#1){\ensuremath{\mathrm{SU}(#1)}}
\def\PSU(#1){\ensuremath{\mathrm{PSU}(#1)}}
\def\SUs(#1){\ensuremath{\mathrm{SU}^*\kern-1pt(#1)}}
\def\U(#1){\ensuremath{\mathrm{U}(#1)}}
\def\SO(#1){\ensuremath{\mathrm{SO}(#1)}}
\def\PSO(#1){\ensuremath{\mathrm{PSO}(#1)}}
\def\SOs(#1){\ensuremath{\mathrm{SO}^*\kern-1pt(#1)}}
\def\SOp(#1){\ensuremath{\mathrm{SO}^+\kern-1pt(#1)}}
\def\O(#1){\ensuremath{\mathrm{O}(#1)}}
\def\Op(#1){\ensuremath{\mathrm{O}^+\kern-1pt(#1)}}
\def\CSO(#1){\ensuremath{\mathrm{CSO}(#1)}}
\def\CSOs(#1){\ensuremath{\mathrm{CSO}^*\kern-1pt(#1)}}
\def\ISO(#1){\ensuremath{\mathrm{ISO}(#1)}}

\def\fsl(#1){\ensuremath{\mathfrak{sl}(#1)}}
\def\su(#1){\ensuremath{\mathfrak{su}(#1)}}
\def\sus(#1){\ensuremath{\mathfrak{su}^*\!(#1)}}
\def\sp(#1){\ensuremath{\mathfrak{sp}(#1)}}
\def\so(#1){\ensuremath{\mathfrak{so}(#1)}}
\def\sos(#1){\ensuremath{\mathfrak{so}^*\!(#1)}}
\def\cso(#1){\ensuremath{\mathfrak{cso}(#1)}}
\def\csos(#1){\ensuremath{\mathfrak{cso}^*\!(#1)}}
\def\iso(#1){\ensuremath{\mathfrak{iso}(#1)}}

\title{Generalised Scherk-Schwarz reductions from gauged supergravity}
\author{Gianluca Inverso}

\begin{document}

\begin{titlepage}

\vspace*{8ex}

\begin{center}
\vskip 1.0cm
{\LARGE \bf Generalised Scherk--Schwarz reductions\\[1ex] from gauged supergravity}\\[3ex]
\vskip 1.0cm

\vspace*{5ex}

{Gianluca Inverso}
\vskip .6cm
{\it
Center for Mathematical Analysis, Geometry and Dynamical Systems, \\
Department of Mathematics, Instituto Superior T\'ecnico,\\
Universidade de Lisboa, Av. Rovisco Pais, 1049-001 Lisboa, Portugal}\\[2ex]
ginverso@math.tecnico.ulisboa.pt

\vskip 1cm

\vskip 1cm
{\bf Abstract}
\end{center}


\begin{narrower}

\noindent
A procedure is described to construct generalised Scherk--Schwarz uplifts of gauged supergravities. 
The internal manifold, fluxes, and consistent truncation Ansatz are all derived from the embedding tensor of the lower-dimensional theory.
We first describe the procedure to construct generalised Leibniz parallelisable spaces where the vector components of the frame are embedded in the adjoint representation of the gauge group, as specified by the embedding tensor.
This allows us to recover the generalised Scherk--Schwarz reductions known in the literature and to prove a no-go result for the uplift of $\omega$-deformed \SO(p,q) gauged maximal supergravities.
We then extend the construction to arbitrary generalised Leibniz parallelisable spaces, which turn out to be torus fibrations over manifolds in the class above.
\end{narrower}

\end{titlepage}
\tableofcontents

\section{Introduction and summary}\label{intro}

There has been much recent activity in the study and construction of new consistent Kaluza--Klein truncations of supergravity theories \cite{Lee:2014mla,Hohm:2014qga,Baron:2014bya,Guarino:2015jca,Guarino:2015qaa,Guarino:2015vca,Inverso:2016eet}.
Such truncations allow to identify a subsector of the configuration space of a theory compactified on an internal manifold, such that the dynamics are encoded in a lower-dimensional gauged supergravity and any solutions of the latter  lift to solutions of its higher-dimensional parent.

We will focus on consistent truncations that preserve as many supersymmetries as the original theory.%
\footnote{For recent results on consistent truncations to less supersymmetric theories see \cite{Malek:2016bpu,Malek:2016vsh,Malek:2017njj}.}
The problem of identifying such truncations is highly nontrivial.
Until recently, the only known class of internal spaces that allow a systematic construction of consistent truncations have been group manifolds. 
Expanding the supergravity fields in terms of left invariant forms on a Lie group guarantees consistency of the truncation by symmetry arguments \cite{Scherk:1978ta,Scherk:1979zr}.
The proof \cite{deWit:1986oxb} that eleven dimensional supergravity on a seven-sphere admits a consistent trucation to \SO(8) gauged maximal supergravity \cite{deWit:1981sst,deWit:1982bul} relied on much more non-trivial techniques \cite{deWit:1986mz} that can be seen as a precursor of the modern generalised and exceptional generalised geometries (EGG) \cite{Gualtieri:2003dx,Hitchin:2004ut,Coimbra:2011nw,Coimbra:2011ky,Coimbra:2012af}.\footnote{See \cite{Nicolai:2011cy} for extra recent results on the $S^7$ truncation, \cite{Ciceri:2014wya} for a similar rewriting of type IIB supergravity, \cite{Baron:2014bya} for hyperboloid reductions based on the same techniques, and \cite{Guarino:2015jca,Guarino:2015vca} for type IIA supergravity on a six-sphere.
Also see \cite{Nastase:1999cb,Nastase:1999kf,Cvetic:2000ah} for other sphere reductions.}
In fact, it is thanks to the recently developed frameworks of EGG and the closely related extended/exceptional field theories (ExFT)\cite{Berman:2010is,Berman:2012uy,Berman:2012vc,Hohm:2013pua,Hohm:2013vpa,Hohm:2013uia,Hohm:2014fxa,Godazgar:2014nqa,Baguet:2016jph}
that we now understand consistent truncations on spheres systematically \cite{Lee:2014mla} in terms of \emph{generalised} Scherk--Schwarz reductions (see also \cite{Aldazabal:2011nj,Geissbuhler:2011mx,Grana:2012rr,Aldazabal:2013mya} for earlier work).
These formalisms allow to repackage the field content of a supergravity theory in order to give a geometrical interpretation to their gauge symmetries and dualities.
The long-sought proof of consistency of the truncation of type IIB supergravity on $S^5$ to \SO(6) gauged maximal supergravity in five dimensions relied on the ExFT framework \cite{Hohm:2014qga}, and consistent truncations on spheres, hyperboloids, twisted tori and products thereof are now well-understood \cite{Hohm:2014qga,Inverso:2016eet}.
Results concerning sphere reductions of massive IIA supergravity \cite{Guarino:2015jca,Guarino:2015qaa,Guarino:2015vca} have also been rephrased in terms of ExFT and EGG \cite{Ciceri:2016dmd,Cassani:2016ncu}.

The inverse problem of identifying which gauged supergravity theories admit an uplift to ten and eleven dimensional supergravities is equally interesting and non-trivial.
Gauged supergravities have an intricate phenomenology of vacua, (super)symmetry breaking patterns, black holes, branes, and domain wall solutions and identifying which models and solutions are embedded in string/M-theory is important.
The modern framework to describe gauged supergravities is the embedding tensor formalism \cite{Nicolai:2000sc,deWit:2004nw,deWit:2005ub,deWit:2007kvg} (see \cite{Samtleben:2008pe,Trigiante:2016mnt} for reviews and further references). 
In this formalism the gauge group and all gauge couplings are specified by an object $`\Theta_A^{\hat\alpha}`$ transforming in a specific representation of the global symmetry group $\cG\times\bbR^+$ of the ungauged theory.
Consistency of the resulting gauged supergravity is encoded into a set of algebraic constraints on the embedding tensor.
It is natural to expect that the requirements for a gauged supergravity to admit an higher dimensional uplift should be phrased in terms of additional constraints on the embedding tensor.

Important examples of  gauged supergravities not admitting a geometric uplift are the $\omega$-deformed \SO(8) gauged supergravities of \cite{DallAgata:2012bb}.
Attempts to find an origin in eleven-dimensional supergravity for these gaugings found an obstruction \cite{deWit:2013ija} and a no-go result was proven later \cite{Lee:2015xga}. 
There is a large class of gaugings descending from the $\omega$-deformed \SO(8) ones by analytic continuation and contraction \cite{DallAgata:2011aa,DallAgata:2012plb,DallAgata:2014tph}.
These are $\omega$-deformed \SO(p,q), \CSO(p,q,r), and `dyonic CSO' gaugings.
All these gaugings have been uplifted \cite{Hohm:2014qga,Guarino:2015jca,Guarino:2015qaa,Guarino:2015vca,Inverso:2016eet} except for the $\omega$-deformed \SO(p,q) models, for which however the no-go result of \cite{Lee:2015xga} does not apply.
It is therefore an open problem to prove whether these models admit a geometric uplift or not.

In (exceptional) generalised geometry as well as in doubled  \cite{Siegel:1993th,Siegel:1993xq,Hohm:2010jy,Hohm:2010pp,Hohm:2010xe} and extended field theory the diffeomorphisms and gauge symmetries along the internal manifold are packaged in terms of a generalised Lie derivative $\bbL$ with parameters living in an extended tangent space and comprising infinitesimal generators for the internal diffeomorphisms and $p$-form gauge transformations.
Fields are repackaged to fill out representations of the duality group $\cG\times\bbR^+$ that becomes the global symmetry of the lower-dimensional supergravity theory if the compactification space is taken to be a standard torus.
The main difference between the two formalisms is that in DFT/ExFT the coordinates of the internal space are formally extended to cover a full representation of $\cG\times\bbR^+$ and a section (or strong) constraint is imposed to determine which of these coordinates are physical.
Upon solution of the constraint one recovers a standard supergravity theory written in terms of an appropriate generalised geometry.

Generalised Scherk--Schwarz reductions are obtained by expanding the supergravity fields in terms of a generalised Leibniz parallelisation, namely a global frame $\hat E_A$ for the generalised tangent bundle satisfying
\begin{formula}\label{leibnintro}
\bbL_{\hat E_A}\hat E_B = -`X_AB^C` \hat E_C\,,
\end{formula}
where $`X_AB^C`$ are constants.
The expansion coefficients are allowed to depend only on the $D$ dimensional external spacetime coordinates and not on the internal ones. 
Their equations of motion\footnote{%
We will always refer to consistent truncations of the classical equations of motion. Avoiding reference to an action principle allows us to include trombone gaugings \cite{LeDiffon:2008sh,LeDiffon:2011wt}.}
become those of a gauged supergravity  where $`X_AB^C`$ is identified with the embedding tensor.\footnote{Strictly speaking we should restrict ourselves to theories with sixteen supercharges or more for which $`X_AB^C`$ encodes the same information as the embedding tensor $`\Theta_A^{\hat \alpha}`$.
This is the lowest amount of supersymmetry allowed in ten dimensions.
For theories with less supercharges our setup is still correct as long as no matter (e.g. hypermultiplet) symmetries are gauged. 
We will refer to both $`X_AB^C`$ and $`\Theta_M^{\hat\alpha}`$ as the embedding tensor.}
We can extend \eqref{leibnintro} to include deformations of the structure of the generalised tangent bundle and Lie derivative induced by massive and/or gauged deformations of the underlying supergravity theory (e.g. the Romans mass deformation of type IIA supergravity \cite{Ciceri:2016dmd,Cassani:2016ncu}).
The more general condition for a Leibniz parallelisation becomes
\begin{formula}\label{leibnintro2}
\bbL_{\hat E_A}\hat E_B +\hat F^0(\hat E_A)\hat E_B= -`X_AB^C` \hat E_C\,.
\end{formula}
where $\hat F^0(\hat E_A)$ is a linear non-derivative operator encoding the massive/gauged deformation of the target higher-dimensional theory.
The general constraints for consistency of such deformations were analysed in \cite{Ciceri:2016dmd}.

It is clear that the problem of finding what spaces admit a generalised Leibniz parallelisation and for what values of $`X_AB^C`$ is much more complicated in generalised geometries and ExFT's than the analogous problem is in standard differential geometry, where any consistent Lie algebra structure constants define a parallelisable manifold which is (a global form of) the associated Lie group.
So far there has been no general procedure to construct generalised Leibniz parallelisations.
Some progress in this direction has recently been made, reproducing some results specific to
DFT \cite{Hassler:2016srl} and to 
four-manifold reductions of \SL(5) ExFT \cite{duBosque:2017dfc}.
In this paper we solve the problem entirely, by taking a `bottom-up' approach.
The procedure we derive applies to double, exceptional and any other extended field theories, including mass-deformed and gauged ones \cite{Ciceri:2016dmd} (see also \cite{Grana:2012rr}), as long as their generalised Lie derivative closes without the need for constrained gauge parameters (i.e., we do not include $\mathrm E_{8(8)}$~ExFT~\mbox{\cite{Hohm:2014fxa})}.
We provide necessary and sufficient conditions for an embedding tensor to give rise to a generalised Leibniz parallelisation \eqref{leibnintro2}, determining the internal space as well as $\hat E_A$ and $\hat F^0$ in the process.
This defines a consistent generalised Scherk--Schwarz uplift of the associated gauged supergravity.
We do so by first showing when and how one can solve \eqref{leibnintro2} locally on a coordinate patch, and then providing a global extension.
%
An advantage of our approach is that because we start locally, we can employ the formalism of ExFT to capture many generalised geometries at once.
The choice of solution of the section constraint (and hence ultimately the higher dimensional theory to which we uplift our gauged supergravity) is dictated by $`X_AB^C`$ itself.
Instead of looking for solutions of \eqref{leibnintro2} for a given EGG, we span an entire class of theories at once.

\bigskip

We summarise here the structure and main results of this paper.
In section~\ref{review} we review some basic aspects of the embedding tensor formalism and of ExFT/EGG, deriving some useful properties of the torsion of a generalised frame as well as the general consistent conditions for flux, massive and gauged deformations of the generalised Lie derivative, extending the recent analysis of \cite{Ciceri:2016dmd}.

The embedding tensor appearing on the right hand side of \eqref{leibnintro2} can be rewritten as $`X_AB^C`=`\Theta_A^{\hat\alpha}``t_{\hat\alpha}\,B^C`$ where $`t_{\hat\alpha}\,B^C`$ generate $\cG\times\bbR^+$.
It defines the gauging of a subgroup $\Gg\subset\cG\times\bbR^+$.
In section~\ref{main}
we provide the most general solution of \eqref{leibnintro2}, focussing first on the subclass of parallelisations in which the non-vanishing vector components of $\hat E_A$ are valued in the adjoint representation of \Gg  specified by $`\Theta_A^{\hat\alpha}`$.
We later lift this restriction, although it is worth noticing that, to this date, the known examples of generalised Leibiniz parallelisations belong to this restricted class.
There are some requirements for generalised Leibniz parallelisation to exist, and we begin with a local analysis.
Focussing on the restricted class just described, in order to solve \eqref{leibnintro2} we must search for a subgroup $\Hg\subset\Gg$ such that the projection $`\Theta_A^{\_m }`$ of the embedding tensor on the $\H\backslash\G$ coset space generators $t_{\_m }$ (at least for some choice of the latter) satisfies the section constraint
\begin{formula}\label{theta SC intro}
`Y^AB_CD``\Theta_A^{\_m }``\Theta_B^{\_n }` = 0\,,
\end{formula}
where the $\cG\times\bbR^+$ invariant tensor $`Y^AB_CD`$ appears in the generalised Lie derivative.
A second, linear constraint on $`X_AB^C`$ (see \eqref{C-like constr for X}) might also be required depending on the theory and specific gauging.
It can be avoided for exceptional field theories if $`X_AB^C`$ does not gauge the trombone symmetry of the higher-dimensional supergravity.
The coset space $\H\backslash\G$ will be (part of) the internal manifold.
Out of the local data on the coset space we construct a local frame $`E_A^M`$ satisfying
\begin{formula}\label{localLeibnIntro1}
\bbL_{E_A}`E_B^M`-`E_A^P``E_B^Q``F_PQ^M` = -`X_AB^C``E_C^M`\,,
\end{formula}
where $`F_PQ^M`$ is a local version of $\hat F^0$, but can also encode background $p$-form field strengths and twists by global symmetries (including trombone scalings) of the higher-dimensional theory.
The frame $`E_A^M`$ can also encode similar contributions, so that the final background is only obtained combining the two objects.
The proof that $`F_PQ^M`$ satisfies all necessary consistency conditions so that our solution of \eqref{localLeibnIntro1} is locally equivalent to a solution of \eqref{leibnintro2} is one of the main results of this paper.
The objects in \eqref{localLeibnIntro1} do not necessarily extend correctly to a global frame and global fluxes, but it is  possible to construct the latter out of $E`{}_A^M`$  and $`F_MN^P`$ if some conditions are met.
The deformations $\hat F^0$ also determine whether the internal space can be extended with extra flat directions, becoming $\cM_{\rm internal} = \H\backslash\G \times \bbT^n$.

Lifting the requirement that the vector components of the frame only sit in the adjoint representation of G, we find that only minor modifications to our procedure must be implemented, which is taken care of in section~\ref{extension}.
In particular, the most general parallelisable space turns out to be a torus fibration over a coset space $\H\backslash\G$
\begin{formula}\label{intspace}
\cM_{\rm internal} \underset{\mathrm{loc.}}{\simeq} \H\backslash\G \times \bbT^n\,.
\end{formula}
where $\H\backslash\G$, by itself, belongs to the class of generalised Leibniz parallelisable spaces described above and the fiber is determined by a central extension of the gauge algebra, again entirely dictated by the embedding tensor.

In section~\ref{examples} we discuss several examples. 
We start by showing that our procedure reproduces standard group manifold reductions as a special case, as well as the consistent Pauli reductions of \cite{Baguet:2015iou}.
Then we move to the uplifts of (maximal) supergravities with gaugings of \SO(p,q) and \CSO(p,q,r) groups.
We focus on the four-dimensional case where there is a very rich structure for such gaugings \cite{DallAgata:2011aa,DallAgata:2012bb,DallAgata:2014tph}.
In particular, we prove a no-go result for the uplift of the non-compact versions of the $\omega$-deformed \SO(p,q) gaugings discussed in \cite{DallAgata:2012bb,DallAgata:2014tph}, extending the previous negative result of \cite{deWit:2013ija} and the no-go of \cite{Lee:2014mla}, which only applied to the compact gauging \SO(8).
We also find that all electric \CSO(p,q,r) gaugings with $r\neq0$ admit an uplift on $\bbT^{p+q}$ with a locally geometric flux which is analogous to the $Q$-flux of the NSNS string.
Our procedure identifies this uplift in terms of a globally defined frame on $\bbR^n$.

We stress that $\H\backslash\G$ need not be a compact manifold, although we can certainly choose to impose such a restriction.
If the coset space is non-compact, we might want to quotient it by the (free) action of some discrete group $\Gamma\subset\G$.
In general the global frame defined on $\H\backslash\G$ becomes multivalued on the quotient space and the resulting background can at best be interpreted as a U-fold geometry where fields jump by $\cG\times\bbR^+$ transformations along the internal space.
Simple examples of such situation are the locally geometric $Q$-flux in type II and heterotic supergravity and the more general examples discussed in section~\ref{examples}.
We make a few extra comments on this point in section~\ref{outro}, where we conclude.

\section{Some prerequisites}\label{review}
\subsection{Gauged supergravities}\label{gsugra}

We denote the global symmetries of the lower-dimensional supergravity theory---the one we want to gauge and uplift---as $\cG\times\bbR^+$, the second factor being the trombone symmetry.\footnote{%
Every supergravity theory has one global trombone symmetry $\bbR^+$ acting as a rescaling of all fields including the metric\cite{Cremmer:1997xj}.}
These symmetries can be gauged by promoting to local a subgroup $\Gg\subset\cG\times\bbR^+$ and using the vector fields $A_\mu^A$ of the theory to construct the gauge connection.
Schematically (ignoring all other covariantisations), the spacetime derivative is covariantised as
\begin{formula}\label{covdecgauged}
\partial_{\mu} \to D_\mu\equiv \partial_{\mu} - A_\mu^A X_A\,,
\end{formula}
where $X_A$ are the generators of the gauge group.
Because there are usually more vectors that gauge generators, $X_A$ may form a redundant basis and/or have vanishing entries.
It is entirely specified by an embedding tensor $`\Theta_A^{\hat\alpha}`$ as
\begin{formula}\label{def emb tens X}
`X_A`\equiv `\Theta_A^{\hat\alpha}`t_{\hat\alpha}\,,\quad
\hat\alpha=0,1,\ldots,\mathrm{dim}\cG\,,
\end{formula}
where $t_{\hat\alpha}$ generate $\cG\times\bbR^+$.

Closure of the gauge algebra is guaranteed by a quadratic constraint
\begin{formula}\label{QC}
[`X_A`,\,`X_B`]=-`X_AB^C``X_C`\,,
\end{formula}
where $`X_AB^C`\equiv`\Theta_A^{\hat\alpha}``t_{\hat\alpha}_\,B^C`$ are the gauge group generators in the $\*R _{\rm v}$ representation of $\cG\times\bbR^+$, which is the (conjugate of the) one in which the vector fields transform.
We will often refer to $`X_AB^C`$ itself as the embedding tensor.
Crucially for our purpose, \eqref{QC} determines that $`X_AB^C`$ can be seen as structure constants of a Leibniz algebra.
This is more general than a Lie algebra, as $`X_{(AB)}^C`$ need not vanish and correspondingly, the standard Jacobi identity of Lie algebras is not satisfied by $`X_AB^C`$.

A general embedding tensor transforms in the $\cG$ representation
\begin{formula}\label{repr X step 1}
`\Theta_A^{\hat\alpha}` \in \*R _{\rm v} \otimes ( \*adj + \*1 )\,,
\end{formula}
where the singlet corresponds to the trombone component just described.
Consistency (supersymmetry and counting of degrees of freedom) of the gauged supergravity restricts the non-trombone components of the embedding tensor to a subset of the irreps contained in the tensor product $\*R _{\rm v} \otimes \*adj $:
\begin{formula}\label{repr X step 2}
`\Theta_A^{\hat\alpha}` \in \*R _{\Theta} + \*R _{\rm v} \subset \*R _{\rm v} \otimes ( \*adj + \*1 )\,.
\end{formula}
The relevant representations are exemplified in Table~\ref{tab:maxreps} for the case of gauged maximal supergravities.

\begin{table}[hbt]
\centering
\begin{tabular}{|c|c|c|c|c|c|c|}
\hline
  $D$  & 9&8&7&6&5&4  \\
\hline
  $\cG$   & $\SL(2)\times\bbR^+$ & $\SL(2)\times\SL(3)$ & $\SL(5)$ & $\SO(5,5)$ & $\ \Esix\ $ & $\ \Eseven\ $  \\
 $\*R _{\rm v}$ & $\*2 _3 +\*1 _{-4}$  & $(\*2 ,\,\*3 ')$ & $\*10 '$ & $\*16 _c$ & $\*27 $ & $\*56 $   \\
 $\*R _\Theta$ & $\*2 _{-3} +\*3 _{4}$ & $(\*2 ,\,\*3 )+(\*2 ,\,\*6 ')$& $\*15 +\*40 '$& $\*144 _c$ & $\*351 '$& $\*912 $  \\
 $\*adj $ & $\*3 +\*1 $ & $(\*3 ,\,\*1 )+(\*1 ,\,\*8 )$ & $\*24 $ & $\*45 $ & $\*78 $ & $\*133 $\\
\hline
\end{tabular}
\caption{Relevant representations for the duality groups of the maximal supergravities.}
\label{tab:maxreps}
\end{table}

\subsection{Generalised geometries and extended field theories}\label{exft}

Exceptional and extended field theories (ExFT) can be seen as a generalisation of the ideas of double field theory (DFT) \cite{Siegel:1993th,Siegel:1993xq,Hohm:2010jy,Hohm:2010pp,Hohm:2010xe}, and are related to (exceptional) generalised geometry (EGG) \cite{Gualtieri:2003dx,Hitchin:2004ut,Coimbra:2011nw,Coimbra:2011ky,Coimbra:2012af} in a way similar to how DFT is related to complex generalised geometry.
The bosonic sector of ExFT looks similar to a gauged supergravity (usually maximal or half-maximal), with a metric, $p$-form fields, and scalar fields parameterising a coset space $\cG/\cH$ ($\cH$ being the maximal compact subgroup),  all living on a $D$-dimensional `external' spacetime but also formally carrying dependence on an extended set of internal coordinates $Y^M$ filling the $\*R _{\rm v}$ representation of $\cG$.
All fields transform covariantly under the duality group $\cG\times\bbR^+$.
Internal gauge symmetries are, instead of a Lie group as for gauged supergravity, an infinite set of transformations called generalised diffeomorphisms acting on covariant fields via a generalised Lie derivative $\bbL$.
The structure and dynamics of the theory are essentially fixed by enforcing invariance under the internal  symmetries and $Y$-dependent diffeomorphisms on the external spacetime \cite{Hohm:2013nja,Hohm:2013pua,Hohm:2013vpa,Hohm:2013uia,Hohm:2014fxa}.
Consistency of the generalised diffeomorphisms will reduce the dependence on $Y^M$  of fields and gauge parameters to only a subset $y^m$ of physical internal coordinates, with $m=1,...,d$.
The resulting theory is a rewriting of a supergravity theory in $D+d$ dimensions where fields are re-packaged in terms of an EGG defined on the internal $d$ dimensional space.
The generalised diffeomorphisms encode all the local symmetry transformation of the supergravity theory with parameters living on the internal space.
If all the dependence on internal coordinates is removed, ExFT reduce to $D$-dimensional ungauged supergravities with global symmetry group $\cG\times\bbR^+$.
We will only be concerned with the structure of the internal gauge symmetries of ExFT.

The generalised Lie derivative can be defined by its action  on a generalised vector $V^M$, $M$ being an index in the $\*R _{\rm v}$ representation of the duality group, as
\begin{formula}\label{genLie}
\bbL_{\Lambda}V^M &\equiv
\Lambda^N\partial_NV^M-V^N\partial_N\Lambda^M+`Y^MP_QN`\partial_P\Lambda^QV^N+(\lambda-\omega)\partial_N\Lambda^N V^M\\[1ex]
&= \Lambda^N\partial_NV^M +\alpha`\bbP_N^M_P^Q`\partial_P\Lambda^QV^N+\lambda\partial_N\Lambda^N V^M\,.
\end{formula}
where $`\bbP_N^M_P^Q`$ is the projector on the Lie algebra of $\cG$, $\alpha$ is a constant which depends on the specific duality group, and $\omega$ is a characteristic weight, also dependent on the specific theory.
All vectors we will be dealing with have density weight $\lambda=\omega$.
The relation between the projector and the invariant tensor $`Y^MN_PQ`$ is
\begin{formula}\label{YtoP}
`Y^MN_PQ` = \delta^M_P\delta_Q^N+\omega\delta^N_P\delta_Q^M -\alpha`\bbP_N^M_P^Q`\,.
\end{formula}

Closure and the Jacobi identity of the generalised Lie derivative can be rewritten as
\begin{align}
\label{closure and jac for L}
[\bbL_{\Lambda},\,\bbL_{\Sigma}] \Gamma^M 
-\bbL_{ [\Lambda , \Sigma] } \Gamma^M  =0\,,\qquad
\bbL_{ \{\Lambda , \Sigma\} } \Gamma^M = 0 \,.
\end{align}
Strictly speaking, the second condition is not the Jacobi identity itself, but implies it \cite{Berman:2012vc}.
The brackets are defined as
\begin{formula}\label{brackets}
[\Lambda ,\,\Sigma]\equiv \frac12(\bbL_\Lambda\Sigma-\bbL_\Sigma\Lambda)\,,\qquad
\{\Lambda ,\,\Sigma\}\equiv \frac12(\bbL_\Lambda\Sigma+\bbL_\Sigma\Lambda)\,.
\end{formula}
Requiring \eqref{closure and jac for L} to hold for arbitrary parameters $\Lambda ,\,\Sigma$ and $\Gamma$ restricts the dimensionality of the internal space according to the following constraints \cite{Berman:2012vc}
\begin{subequations}
\label{SC}
\begin{align}
\label{SCa}
`Y^MN_PQ`\partial_M\partial_N = 0\,,\\
\label{SCb}
(
`Y^MP\!_RS`\delta_N^Q - `Y^MP\!_TN``Y^TQ\!_RS`
) `\partial_(P``\partial_Q)` =0\,,\\
\label{SCc}
(
`Y^MP\!_TN``Y^TQ\!_[SR]`+2`Y^MP\!_[R|T``Y^TQ\!_S]N`-`Y^MP\!_[RS]`\delta^Q_N-2`Y^MP\!_[S|N`\delta_{R]}^Q
) `\partial_(P``\partial_Q)` =0\,,\\
\label{SCd}
(
`Y^MP\!_TN``Y^TQ\!_(SR)`+2`Y^MP\!_(R|T``Y^TQ\!_S)N`-`Y^MP\!_(RS)`\delta^Q_N-2`Y^MP\!_(S|N`\delta_{R)}^Q
) `\partial_[P``\partial_Q]` =0\,.
\end{align}
\end{subequations}
In all ExFT's discussed so far in the literature, \eqref{SCb}--\eqref{SCd} are implied by \eqref{SCa}, which is referred to as the section constraint \cite{Berman:2012vc}.
The two derivatives can act either on the same field (or products of fields) or on different ones (or products), which motivates the symmetrisations.
This means that \eqref{SC} must be solved algebraically by writing
\begin{formula}\label{secconrepres}
\partial_M \equiv `\cE_M^m`\partial_m\,,
\end{formula}
where $`\cE_M^m`$ is a constant rectangular matrix of maximal rank satisfying the constraints above, and $\partial_m$ are the physical internal derivatives.
Clearly there will be upper bounds to the dimensionality of the internal space.
We will always assume that the section constraint is satisfied and
will often leave contraction with $`\cE_M^m`$ as understood, writing for example $\Lambda^m\equiv \Lambda^M`\cE_M^m`$, so that the section constraint becomes $`Y^mn_PQ`=0$.
Two choices of $`\cE_M^m`$ are equivalent if they are related by $\cG\times\bbR^+$ acting on the $\*R _{\rm v}$ index.
Each inequivalent solution of the section constraint determines an EGG on the internal space with coordinates $y^m$ and derivatives $\partial_m$.
For instance, the maximal solutions of the section constraint in exceptional field theory reproduce the series of EGG's of eleven-dimensional supergravity and type IIB supergravity \cite{Hohm:2013pua,Hohm:2013vpa,Hohm:2013uia}.

Once we have fixed our choice of $`\cE_M^m`$ we can identify a few important subgroups of $\cG\times\bbR^+$.
First, there is a subgroup \GL(d) such that $m$ corresponds to the fundamental index and
\begin{formula}\label{GLd def from E}
`g_M^N``\cE_N^n``g-1_n^m` = `\cE_M^m`\,,\qquad g\in\GL(d)\subset\cG\times\bbR^+\,.
\end{formula}
This is identified with the (standard) structure group of the internal manifold.
Second, there is a subgroup $(\cG_0\times\bbR^+_0)\ltimes\cP_0$ where $\cG_0\times\bbR^+_0$ commutes with \GL(d) and $\cP_0$ is generated by a nilpotent algebra, such that
\begin{formula}\label{G0 and P0 def from E}
`U_M^N``\cE_N^m` = `\cE_M^m` \,,\qquad `U_M^N`\in(\cG_0\times\bbR^+_0)\ltimes\cP_0\subset\cG\times\bbR^+\,.
\end{formula}
The unipotent group $\cP_0$ corresponds to shifts of the $p$-form potentials of the higher-dimensional supergravity theory and completes \GL(d) to the (split) generalised structure group of the generalised tangent bundle
\begin{formula}\label{S0}
\GL(d)\ltimes\cP_0\,.
\end{formula}
Transition functions on the generalised tangent bundle take values in this group.
The $\cG_0\times\bbR^+_0$ group corresponds to internal global symmetries for the higher dimensional theory, $\bbR^+_0$ being its trombone symmetry. If the higher dimensional theory is gauged and/or massive, these are the global symmetries of its ungauged, massless sibling.

The consistent truncation of a supergravity theory living in $D+d$ dimensions down to a $D$-dimensional gauged supergravity with the same amount of supersymmetries is obtained by identifying a frame $\hat E_A^M(y)$ on the internal manifold satisfying the Leibniz parallelisation condition \eqref{leibnintro2}.
Then, all the $y^m$ dependence of the fields is factorised in terms of $\hat E_A^M(y)$ and $y$-independent coefficient fields that will become the gauged supergravity fields.
A thorough discussion of this factorisation process, taking into account the truncation of the tensor hierarchy associated with the internal gauge structure is carried out in \cite{Hohm:2014qga}.

\subsection{Torsion induced by a generalised frame}

Let us now introduce the torsion associated with a frame $`E_A^M`$ and summarise some of its properties.
As a matrix, the frame is an element of $\cG\times\bbR^+$.
Its torsion $`T_AB^C`$ can be defined as
\begin{formula}\label{torsion from Lie der}
\bbL_{E_A}`E_B^M` \equiv -`T_AB^C``E_C^M` \,,
\end{formula}
and is usually $y^m$-dependent.
The indices $A,B,C,...$ are spectators with respect to the Lie derivative.
A more explicit expression is written in terms of the (generalised) Weitzenb\"ock connection coefficients
\begin{align}
\label{weitzenbock}
`W_AB^C`&\equiv `E_A^m``E_B^N`\partial_m`E_N^C` \,,\\[1ex]
\label{torsion from weitz}
`T_AB^C` &\equiv 2`W_[AB]^C`+`Y^CD_EB``W_DA^E`\,.
\end{align}
Because is by definition invariant under $\cG\times\bbR^+$, we can also write it with spectator indices as done above.
The torsion sits in the same $\cG\times\bbR^+$ representations $\*R _\Theta+\*R _{\rm v}$ as the embedding tensor of gauged supergravity.

We now consider the generalised Lie derivative of two objects $\Lambda^A`E_A^M`$, $\Sigma^A`E_A^M`$, where both $\Lambda^A$ and $\Sigma^A$ are $y^m$-dependent and arbitrary.
Their generalised Lie derivative can be written as
\begin{formula}\label{genLie with frame torsion}
&\bbL_{\Lambda^A E_A}(\Sigma^C`E_C^M`) =\\[1ex]
&\hfill=\left(\Lambda^A`E_A^m`\partial_m\Sigma^C - \Sigma^A`E_A^m`\partial_m\Lambda^C +`Y^CD_EF``E_D^m`\partial_m\Lambda^E \Sigma^F -\Lambda^A\Sigma^B`T_AB^C`\right)`E_C^M`\,.
\end{formula}
Because all the objects in these expressions satisfy the section constraints, this generalised Lie derivative satisfies the closure and Jacobi relations \eqref{closure and jac for L}:
\begin{align}
\label{closure for torsion}
[\bbL_{\Lambda^A E_A},\,\bbL_{\Sigma^B E_B}] (\Gamma^C`E_C^M`) - \bbL_{ [\Lambda^A E_A, \Sigma^B E_B] } (\Gamma^C`E_C^M`)  &=0\,,\\[1ex]
\label{jac for torsion}
\bbL_{ \{\Lambda^A E_A, \Sigma^B E_B\} } (\Gamma^C`E_C^M`) &= 0 \,.
\end{align}
These expressions imply some useful properties for $`T_AB^C`$.
First, combining \eqref{closure for torsion} and \eqref{jac for torsion} and taking $\Lambda^A$, $\Sigma^B$ and $\Gamma^C$ to be constant, we arrive at an expression that generalises the closure constraint of the embedding tensor to a $y^m$-dependent torsion:\footnote{An analogous computation was performed in \cite{Aldazabal:2013mya}.}
\begin{formula}\label{BI for torsion}
&`T_AC^F``T_BF^D`-`T_BC^F``T_AF^D`+`T_AB^F``T_FC^D`+\\
&\quad\hfill+`E_A^m`\partial_m`T_BC^D`-2`E_[B^m`\partial_m`T_|A|C]^D`
-`Y^DF_GC``E_F^m`\partial_m`T_AB^G` = 0\,.
\end{formula}
Notice that the last two terms correspond to a torsion projection on the indices $BCD$ analogous to \eqref{torsion from weitz}.

Substituting \eqref{BI for torsion} into \eqref{jac for torsion} and taking  $\Sigma^B$ and $\Gamma^C$ constant (but not $\Lambda^A$), we arrive at an expression which is analogous to the C-constraint of \cite{Ciceri:2016dmd}, but with some extra terms:
\begin{formula}\label{C-like for torsion}
&\Big[`T_(CD)^A`\delta_F^B-`Y^AB_HF``T_(CD)^H`-\frac12`Y^HB_CD``T_HF^A`+ 
\\&\hspace{6em}
+\frac12\big(`Y^AI_CD`\delta^H_F-`Y^AH_JF``Y^JI_CD`\big)`W_HI^B`\Big]`E_B^m` = 0 \,.
\end{formula}
This expression is covariant under generalised diffeomorphisms by virtue of \eqref{SCb}.
We stress again that \eqref{BI for torsion} and \eqref{C-like for torsion} are properties automatically satisfied by the torsion of a frame $`E_A^M`$.

\subsection{Deformations of generalised diffeomorphisms}\label{genflux props}

Let us now consider a different situation, expanding on the analysis of \cite{Ciceri:2016dmd}.
We introduce a torsion-like term $`F_MN^P`$, which we dub the generalised flux, in the generalised Lie derivative. 
It also sits in the $\*R _\Theta+\*R _{\rm v}$ representations.
We do \emph{not} assume that it arises from some (local) frame as was the case for the last term in \eqref{genLie with frame torsion}.
We define a deformed generalised Lie derivative $\widetilde\bbL$ as
\begin{formula}\label{Ltilde}
\widetilde\bbL_{\Lambda} \Sigma^M \equiv
\Lambda^m\partial_m \Sigma^M- \Sigma^m\partial_m\Lambda^M + `Y^Mm_PQ`\partial_m \Lambda^P \Sigma^Q - \Lambda^P\Sigma^Q`F_PQ^M` \,.
\end{formula}
Notice that compared to \eqref{genLie with frame torsion}, there are no `spectator' indices here.
Because we have introduced $`F_MN^P`$ by hand, this time we have no guarantee that $\widetilde\bbL$ satisfies closure and Jacobi relations analogous to \eqref{closure and jac for L}.
We must thus \emph{impose}
\begin{align}
\label{closure and jac for F}
[\widetilde\bbL_{\Lambda},\,\widetilde\bbL_{\Sigma}] \Gamma^M 
-\widetilde\bbL_{ [\Lambda , \Sigma]_{\rm F} } \Gamma^M  =0\,,\qquad
\widetilde\bbL_{ \{\Lambda , \Sigma\}_{\rm F} } \Gamma^M = 0 \,.
\end{align}
The new brackets are
\begin{formula}\label{brackets}
[A,\,B]_{\rm F}\equiv \frac12(\widetilde\bbL_AB-\widetilde\bbL_BA)\,,\qquad
\{A,\,B\}_{\rm F}\equiv \frac12(\widetilde\bbL_AB+\widetilde\bbL_BA)\,.
\end{formula}

The resulting constraints that $`F_MN^P`$ must satisfy have been analysed in \cite{Ciceri:2016dmd} assuming constancy of $`F_MN^P`$ and absence of the trombone component: $`F_MP^P`=0$.
Here we directly write the final requirements for a general $`F_MN^P`$ (satisfying the section constraint).
First, the generalised flux must satisfy the `X-' and `C-constraints' of \cite{Ciceri:2016dmd}, which are not equivalent for a generic theory and for non-vanishing trombone components.
These constraints read respectively
\begin{align}
\label{X-constraint}
&`F_MN^P``\cE_P^m` = 0 \,,\\[1ex]
\label{C-constraint}
& C[F]`{}_SPQ^MN``\cE_N^m` \equiv
\Big(`F_(PQ)^M`\delta_S^N -`Y^MN_TS``F_(PQ)^T`-\frac12`Y^TN_PQ``F_TS^M`\Big)`\cE_N^m` = 0 \,.
\end{align}

Second, the generalised flux must satisfy a generalised Bianchi identity not dissimilar to the torsion property \eqref{BI for torsion}:\footnote{This expression was derived with Franz Ciceri and Adolfo Guarino in the making of \cite{Ciceri:2016dmd} and \cite{Ciceri:2016hup}.}
\begin{formula}\label{BI}
&`F_MP^R``F_NR^Q`-`F_NP^R``F_MR^Q`+`F_MN^R``F_RP^Q`+\\
&\quad\hfill+`\cE_M^m`\partial_m`F_NP^Q`-2`\cE_[N^m`\partial_m`F_|M|P]^Q`
-`Y^QR_SP``\cE_R^m`\partial_m`F_MN^S` = 0\,.
\end{formula}
This expression reduced to the embedding tensor closure constraint in the analysis of \cite{Ciceri:2016dmd}.

The constraint \eqref{X-constraint} guarantees in particular that $`F_MN^P`$ does not affect the algebra of standard internal diffeomorphisms, generated by vectors of the schematic form $\Lambda^M = (\Lambda^m,\,0...0)$, i.e. non-vanishing only along the tangent space components.
Thus the generalised flux induces deformations of the internal gauge symmetries of the supergravity associated with the extended generalised geometry.
Such deformations can be due to background $p$-form fluxes, twists of the field content by coordinate-dependent $\cG_0\times\bbR^+_0$ transformations,\footnote{For instance, in type IIB supergravity there is an \SL(2) triplet of 1-form fluxes including the RR $F_1$ and the dilaton flux. They originate from a coordinate dependent \SL(2) twists of the fields of the theory, analogous to the compactifications with duality twists of \cite{Dabholkar:2002sy}.} massive deformations, and gaugings.
The requirements above guarantee that the resulting set of gauge symmetries is consistent.
Indeed, the analysis of \cite{Ciceri:2016dmd} already shows that $`F_MN^P`$ reproduces exactly the standard $p$-form fluxes of 11d and type II supergravities (assuming the respective solutions of the section constraint are adopted), including the Romans mass in type IIA and a triplet of \SL(2) one-form fluxes in type IIB supergravity.
A similar analysis was carried out for \SL(2)-DFT in \cite{Ciceri:2016hup}.
Here we are extending the analysis to non-constant fluxes and also allow for a `trombone flux' arising by coordinate dependent trombone rescalings of the higher-dimensional fields.\footnote{We can regard trombone gauged IIA supergravity \cite{Howe:1997qt} as arising from eleven-dimensional supergravity exactly through such a `trombone flux' compactification on a circle.}

One more useful property of the generalised flux is its Lie derivative.
We assign density weight $-\omega$ to $`F_MN^P`$ for consistency of the deformed Lie derivative $\widetilde\bbL$. 
Using the C-constraint we find \cite{Ciceri:2016dmd}
\begin{formula}\label{Lie der of F}
\bbL_{\Lambda} `F_MN^P` = \Lambda^m\partial_m `F_MN^P` +2`\cE_[M^m`\partial_m \Lambda^T `F_|T|N]^P` + `Y^PS_RN``\cE_S^m`\partial_m\Lambda^T`F_TM^R`\,.
\end{formula}

It should be stressed that most components of $`F_MN^P`$ can be re-absorbed into a twisting of the covariant tensors by some (locally defined) matrix $C(y^m)`{}_M^N`$ satisfying
\begin{formula}\label{E-constraint}
`C_M^N``\cE_N^m`=`\cE_M^m` \,,
\end{formula}
so that the induced flux is the torsion projection \eqref{torsion from weitz} of $`\cE_M^m`\partial_m`C_N^Q``C-1_Q^P`$ and it satisfies \eqref{X-constraint}, \eqref{C-constraint} and \eqref{BI} automatically.
The matrix $`C_M^N`$ will be determined by the $p$-form potentials associated with fluxes in $`F_MN^P`$, and thus be only defined patch-by-patch in the internal space.
This is analogous to the twisting procedure discussed in \cite{Coimbra:2011nw,Coimbra:2011ky,Coimbra:2012af} to locally map the generalised tangent bundle into  global vectors and $p$-forms, although we must stress that in the current local setup further twistings are allowed, such as those inducing dilaton flux (or the full triplet of \SL(2,\bbR) Scherk--Schwarz flux in Type IIB supergravity), and the one associated with a non-vanishing trombone component.
These extra twists by global symmetries necessarily correspond to 1-form \GL(d) components of $`F_MN^P`$, because the associated $`C_M^N`$ is a \GL(d) singlet.
On the other hand, components of $`F_MN^P`$ taking values in the algebra of $\cG_0\times\bbR^+_0$ and being \GL(d) singlets will not be integrable and will necessarily correspond to embedding tensor components of the higher dimensional theory.%

\section{Generalised Leibniz parallelisations from gauged supergravity}
\label{main}

\subsection{Local uplift}\label{recipe}

We now come to the main part of this paper.
Suppose we have a gauged supergravity with embedding tensor $`X_AB^C`$ satisfying the representation and quadratic constraints.
In order to find an uplift of such theory to a higher dimensional supergravity with the same amount of supersymmetries, we need to find a frame $\hat E`{}_A^M`$ and possibly some non-trivial deformation $\hat F`{}0_MN^P`$ satisfying \eqref{X-constraint}, \eqref{C-constraint} and \eqref{BI} such that the generalised Scherk--Schwarz condition is satisfied:
\begin{formula}\label{SS condition twisted}
\bbL_{\hat E_A}\hat E`{}_B^M` -\hat E`{}_A^P`\hat E`{}_B^Q`\hat F^0`{}_PQ^M`  = -`X_AB^C`\hat E`{}_C^M` \,.
\end{formula}
As we show below, we find it more convenient to allow part of $\hat E`{}_A^M`$ to be absorbed in the generalised flux $`F_MN^P`$, so that one can look for the equivalent requirement
\begin{formula}\label{SS condition}
\widetilde\bbL_{E_A}`E_B^M` = -`X_AB^C``E_C^M` \,,
\qquad \hat E`{}_A^M` \equiv `E_A^N` `C_N^M`\,,
\end{formula}
provided $`F_MN^P`$ satisfies all consistency constraints.%
\footnote{The generalised flux also contains the information originally encoded into $\hat F^0$.}

Let us assume that a generalised frame $\hat E_A$ satisfying \eqref{SS condition twisted} exists for a solution of the section constraint determined by $`\cE_M^m`$.
Then, $\hat E`{}_A^M``\cE_M^m`\equiv `K_A^m`$ are vectors with (standard) Lie bracket
\begin{formula}\label{Liebrack}	
[K_A,\,K_B] = -`X_AB^C` K_C\,.
\end{formula}
A first consequence of \eqref{Liebrack} is that $`X_(AB)^C`K_C=0$.
Exploiting this fact we conclude that projecting either the index $A$ or the index $B$ onto the left kernel of $`\Theta_A^a`$, the right hand side of \eqref{Liebrack} vanishes.
Therefore we can write
\begin{formula}\label{vectoralgebra}
`X_(AB)^C`K_C &=0\,,\qquad
[K_A,\,K_B] =`\Theta_A^a``\Theta_B^b`( `f_ab^c`K_c + `h_ab^{c_0}`K_{c_0})\,,
\end{formula}
where $`h_ab^{c_0}`=`h_[ab]^{c_0}`$ are components of the embedding tensor encoding a central extension of the \Gg algebra (see for instance \cite{deWit:2007kvg}) and $c_0$ runs over entries different than $a,\,b,\,c$.
For simplicity in this and in the next section we will assume that the only non-vanishing vector components of $\hat E_A$ are the \Gg vectors $K_a$, so that $\hat E`{}_A^M``\cE_M^m`=`\Theta_A^a``K_a^m`$.
Once this case is well-understood, the extension of the procedure in presence of central charges  turns out to be relatively straightforward and we discuss it in section~\ref{extension}.

Because $\hat E_A$ is everywhere non-vanishing, this implies that there are always $d$ linearly independent vectors among the $K_a$ at each point on the manifold and therefore we have a homogeneous space $\Hg\backslash\Gg$ with $`K_a`$ generating the transitive action of G on the manifold.
We introduce the coset representatives $L(y)$ of $\H\backslash\G$ with transformation property
\begin{formula}\label{coset repr}
L(y)g = h(y')L(y')\,,\ \ g\in\G\,,\ h(y)\in\H\,.
\end{formula}
Out of the coset representative we can define the Cartan--Maurer form $\Omega$, reference Vielbein $\oe$ and \Hg connection $Q$
\begin{formula}\label{CM form and Vielbein}
\Omega_m \equiv \partial_mLL^{-1} \equiv `\oe_m^{\_m }`t_{\_m }+`Q_m^i`t_i\,,
\end{formula}
where $i$ runs along the algebra of H.
The infinitesimal version of \eqref{coset repr} implies that 
\begin{formula}\label{killonrepr}
`\Theta_A^a``K_a^m`=(LX_AL^{-1})|^{\_m }`\oe_{\_m }^m`
=`L-1_A^B``\Theta_B^{\_m }``\oe_{\_m }^m`
\end{formula}
where $|^{\_m }$ is the projection onto the coset generators and in the last step we have used gauge invariance of the embedding tensor.
We thus conclude that 
\begin{formula}\label{fserbghtafd}
\hat E`{}_A^M``\cE_M^m``\oe_m^{\_m }` = `L{-1}_A^B``\Theta_B^{\_m }`\,.
\end{formula}
Because the left hand side satisfies the section constraint, so does the right hand side, which implies that as a matrix $`\Theta_A^{\_m }`$ can only differ from $`\cE_M^m`$ by a $\cG\times\bbR^+$ transformation (which we can reabsorb in $\hat E_A$) and that it must satisfy the section constraint
\begin{formula}\label{sec constr Theta}
`Y^AB_CD``\Theta_A^{\_m }``\Theta_B^{\_n }` = 0\,.
\end{formula}
This means that we can map the extended internal space derivatives $\partial_M$ into the physical internal derivatives $\partial_m$ as
\begin{formula}\label{internal der from Theta}
\partial_M \equiv `\cE_M^m`\partial_m\,,\qquad `\cE_M^m` = `\delta_M^A``\delta_{\_m }^m``\Theta_A^{\_m }`\,.
\end{formula}
From now on we will simply write $`\Theta_M^m`$ in place of $`\cE_M^m`$.

It is now helpful to notice that as matrices, $`\oe_m^{\_m }`$ and its inverse $`\oe_{\_m }^m`$ are elements of \GL(d) and have a natural embedding into $\cG\times\bbR^+$ which reads
\begin{formula}\label{ref Vielbein in GL}
`\oe_M^A`\,,\ `\oe_A^M`\,\ \in \GL(d)\subset\cG\times\bbR^+\,.
\end{formula}
Notice in particular that \eqref{sec constr Theta} implies 
\begin{formula}\label{vielb block structure}
`\oe_M^A``\Theta_A^{\_m }` = `\Theta_M^m``\oe_m^{\_m }`\,,\qquad
`\oe_A^M``\Theta_M^m` = `\Theta_A^{\_m }``\oe_{\_m }^m`\,.
\end{formula}

At this point we notice that any two candidate expressions for $\hat E`{}_A^M`$ that are equal along the vector components can only differ by terms absorbable into the generalised flux $`F_MN^P`$ through some locally defined matrix $`C_M^N`$, as done going from \eqref{SS condition} to \eqref{SS condition twisted}.
This means that there is no loss of generality in seeking local solutions of \eqref{SS condition twisted} by solving instead \eqref{SS condition} with the Ansatz\footnote{This is similar to the expression provided in \cite{duBosque:2017dfc} for the specific case of consistent truncations from eleven to seven dimensions via \SL(5) ExFT. There should be an exact match when we restrict to their case. However notice that we do not need to introduce the necessary background flux by hand, as it will be automatically generated in our procedure.}
\begin{formula}\label{frame Ansatz}
`E_A^M` \equiv `L-1_A^B``\oe_B^M`\,,
\end{formula}
and the flux will just be the difference between the torsion of $`E_A^M`$ and the embedding tensor, dressed with the frame itself
\begin{formula}\label{F from T and X}
`F_MN^P` = E\circ (X-T)`{}_MN^P`\,,
\end{formula}
where for convenience we will often use the shorthand notation
\begin{formula}\label{shorthand o}
E\circ `X_MN^P` \equiv `E_M^A``E_N^B``X_AB^C``E_C^P`\,.
\end{formula}
For $`E_A^M`$ and $`F_MN^P`$ to extend to globally well-defined objects the definitions above must be amended without affecting the final result \eqref{SS condition}.
This is done in section~\ref{global} and in appendix~\ref{appLdeco}.
It is however more convenient to locally solve \eqref{SS condition} using the definitions above.

One may worry that the definition \eqref{F from T and X} renders \eqref{SS condition} trivial as we are just subtracting the torsion of $`E_A^M`$ from the required result.
This is not so because $`F_MN^P`$ is severely restricted by the consistency conditions \eqref{X-constraint}, \eqref{C-constraint} and \eqref{BI}.
An important part of our work is to prove that \eqref{F from T and X} satisfies these constraints.

Let us make a short summary.
What we have found so far is that any solution to \eqref{SS condition twisted} for a given choice of $`X_AB^C`$ (and a-priori undetermined $\hat F^0$) such that the only non-vanishing vector components of $\hat E_A$ are the \Gg vectors $K_a$, can be locally encoded into a frame $`E_A^M`$ and a flux $`F_MN^P`$ satisfying \eqref{SS condition} and the flux consistency conditions \eqref{X-constraint}, \eqref{C-constraint} and \eqref{BI}.
We have also found that a necessary requirement for such uplifts to exists is that one can choose the coset generators $t_{\_m }$  so that the projection of the embedding tensor onto $t_{\_m }$ satisfies the section constraint \eqref{sec constr Theta}.

In all cases where the `C-'constraint \eqref{C-constraint} is implied by \eqref{X-constraint} the section constraint \eqref{sec constr Theta} is sufficient for consistency of the local solution.
This is the case in particular for double and exceptional field theories, as long as $`F_MN^P`$ does not involve a gauging of the trombone \cite{Ciceri:2016dmd}.\footnote{A counterexample is \SL(2)-DFT \cite{Ciceri:2016hup}.}
We will see below that this can be avoided by requiring that $`X_AB^C`$ does not gauge a certain $\bbR^+_0\subset\cG\times\bbR^+$ corresponding to the trombone symmetry of the higher dimensional theory.
Whenever \eqref{X-constraint} and \eqref{C-constraint} are inequivalent we find that a further linear requirement must be imposed on the embedding tensor:
\begin{formula}\label{C-like constr for X}
C[X]`{}_FCD^AB``\Theta_B^{\_m }`
+\frac14\big(`Y^AI_CD`\delta^H_F-`Y^AH_JF``Y^JI_CD`\big)
  \big(`X_HI^B`+2`\Theta_(H^{\_m }``t_{\_m }I)^B`\big)`\Theta_B^{\_m }` = 0\,,
\end{formula}
where $C[X]$ is defined as in \eqref{C-constraint}.
This is a necessary requirement for consistency of $`F_MN^P`$ as we will show in the proof.
Because $\hat F`{}0_MN^P`$ differs only by terms induced by some $`C_M^N`$, which cannot induce a violation of \eqref{C-constraint}, the requirement \eqref{C-like constr for X} is also necessary for consistency of $\hat F`{}0_MN^P`$.

An important consequence of \eqref{sec constr Theta} is that 
\begin{formula}\label{H in struct x glob}
\H \subset (\GL(d)\times\cG_0\times\bbR^+_0)\ltimes\cP_0\,,
\end{formula}
where $\GL(d)\ltimes\cP_0$ is the generalised structure group on the internal manifold and $\cG_0\times\bbR^+_0$ are the global symmetries of the higher dimensional theory\footnote{In its ungauged, massless flavour.} living in $D+d$ dimensions.
To prove this we take a transformation $h\in\H$ and use gauge invariance of $`\Theta_A^\alpha`$ to write 
$
`h_A^B``\Theta_B^\alpha` = `\Theta_A^\beta``h_\beta^\alpha`\,.
$
Combining this with closure of H we arrive at
\begin{formula}\label{H in S proof step 1}
`h_A^B``\Theta_B^{\_m }` = `\Theta_A^{\_n }``h_{\_n }^{\_m }`\,.
\end{formula}
Mapping this to an action on $\partial_M$, we have
\begin{formula}\label{H in S proof step 2}
`h_M^N` \in \H\,,\qquad
`h_M^N`\partial_N = `h_M^N``\Theta_N^m`\partial_m = `\Theta_M^n``h_n^m`\partial_m
\end{formula}
which means by definition that $`h_M^N`$ acts on $\partial_M$ as a \GL(d) transformation on the physical $\partial_m$, respecting the choice of solution of the section constraint.
The most general transformation with this property is indeed of the type in \eqref{H in struct x glob}.

Another important consequence of our requirements is that $`\Theta_A^{\_m }`$ is automatically \GL(d) invariant, which in turn guarantees consistency of the identification \eqref{internal der from Theta}.
For future use, we also stress that the quadratic constraint \eqref{QC} implies in particular a symmetrised version of \eqref{X-constraint} for $`X_AB^C`$:
\begin{formula}\label{sym X constr for X}
`X_(AB)^C``\Theta_C^{\_m }` = 0\,.
\end{formula}

\subsection{Proof of consistency}\label{proof}

We begin by proving that the flux \eqref{F from T and X} satisfies \eqref{X-constraint}.
This is guaranteed by $E`{}_A^m`=`\Theta_A^a``K_a^m`$ and \eqref{vectoralgebra}, which tells us that 
$
(`T_AB^C`-`X_AB^C`)`E_C^m` = 0\,
$
and in turn, by conjugation with the frame, gives us \eqref{X-constraint}.

It is useful to map this simple proof to a property of the Weitzenb\"ock connection.
Multiplying by $L(y)$ the expression above and using the gauge invariance of $`X_AB^C`$ we arrive at
\begin{formula}\label{vector condition on torsion}
(L\circ `T_AB^C`-`X_AB^C`)`\Theta_C^{\_m }` = 0
\end{formula}
and using \eqref{torsion from weitz} and the section constraint we finally obtain
\begin{formula}\label{rho component of torsion}
L\circ `T_AB^C``\Theta_C^{\_m }` = 2 L\circ `W_[AB]^C``\Theta_C^{\_m }` =`X_AB^C``\Theta_C^{\_m }`\,.
\end{formula}
Notice how consistency of this identity relies on the identification of the solution of the section constraint $`\cE_M^m`$ with $`\Theta_M^m`$, which guarantees antisymmetry of the right hand side, c.f. \eqref{sym X constr for X}.

We must now prove the C-constraint \eqref{C-constraint} for $`F_MN^P`$.
For double and exceptional field theories this requirement is redundant as long as $`F_MN^P`$ does not contain a trombone component \cite{Ciceri:2016dmd}, but it needs to be proven for all other cases.
Substituting \eqref{F from T and X} into \eqref{C-constraint}, recalling $`\cE_M^m`=`\Theta_M^m`$, using \eqref{C-like for torsion} and re-dressing the expression with $`\oe_A^M`$ we arrive at
\begin{formula}\label{wesdtcfgh}
C[X]`{}_FCD^AB``\Theta_B^{\_m }` = -\frac12\big(`Y^AI_CD`\delta^H_F-`Y^AH_JF``Y^JI_CD`\big)L\circ `W_HI^B``\Theta_B^{\_m }`\,.
\end{formula}
The $HI$-antisymmetric part of $L\circ `W_HI^B``\Theta_B^{\_m }`$ is already given in \eqref{rho component of torsion}.
We are only left with evaluating the contribution of the symmetric part.
To this purpose we evaluate the Weitzenb\"ock connection coefficients from \eqref{frame Ansatz} to find
\begin{formula}\label{weitz of coset first version}
L\circ `W_HI^B` 
= `\Theta_H^{\_n }`(\ow_{\_n }+`\oe_{\_n }^m``\Omega_m`)`{}_I^B`
= `\Theta_H^{\_n }`(\ow_{\_n }+t_{\_n }+`\oe_{\_n }^m``Q_m`)`{}_I^B`\,,
\end{formula}
where $`\ow_{\_mn }^{\_p }`\equiv `\oe_{\_m }^m``\oe_{\_n }^n`\partial_m`\oe_{n }^{\_p }`$ appears in \eqref{weitz of coset first version} embedded into the Lie algebra of $\cG\times\bbR^+$ analogously to how we embedded $\oe$ in \eqref{ref Vielbein in GL}.%
\footnote{%
For \GL(d) algebra elements such as $\ow$ an explicit expression for this embedding is 
\begin{formula}\label{gld embedding}
`\ow_{\_m A}^B`\equiv 
`\ow_{\_mn }^{\_p }`\,`\Theta_C^{\_n }`\bar\Theta`{}_{\_p }^D`\big(\delta^C_A\delta_D^B-`Y^BC_DA`\big)\,,
\end{formula}
where $\bar\Theta`{}_{\_m }^A`$ is the unique pseudoinverse of $`\Theta_A^{\_m }`$ such that the projector $`\Theta_A^{\_m }`\bar\Theta`{}_{\_m }^B`$ is orthogonal.
}
Symmetrising in $HI$ and contracting with $`\Theta_B^{\_m }`$ we notice that because  $\ow$ and $Q_m$ take values in the Lie algebra of $(\GL(d)\times\cG_0\times\bbR^+_0)\ltimes\cP_0$, their contributions to \eqref{wesdtcfgh} take the form
\begin{formula}\label{erfgb}
`\Theta_(H^{\_n }``\Theta_I)^{\_p }`\,(`\ow_{\_np }^{\_m }`-`\oe_{\_n }^m``Q_m^i``f_i{\_p }^{\_m }`)\,,
\end{formula}
and thus vanish because of the section constraints.
The remaining contributions from \eqref{rho component of torsion} and \eqref{weitz of coset first version} add up to reproduce the consistency requirement \eqref{C-like constr for X} on the embedding tensor, concluding the proof that $`F_MN^P`$ satisfies the C-constraint \eqref{C-constraint}.

For future reference it is convenient to write an explicit expression for $`F_MN^P`$.
To do so we define the projection onto the algebra of $(\cG_0\times\bbR^+_0)\ltimes\cP_0$ as
\begin{formula}\label{parab projection def}
\breve t_{\hat \alpha} \equiv \bbP_{(\cG_0\times\bbR^+_0)\ltimes\cP_0}(t_{\hat \alpha})
\end{formula}
and similarly on any other object valued in the duality algebra.
Using \eqref{F from T and X}, \eqref{rho component of torsion}, \eqref{weitz of coset first version} and projecting onto $(\cG_0\times\bbR^+_0)\ltimes\cP_0$ we arrive at
\begin{formula}\label{explicit F}
\oe\circ `F_AB^C` =\ &
\breve X`{}_AB^C`
+\frac12`f_{\_mn }^{\_p }`\big(`\Theta_B^{\_m }``Y^C{\_n }_{\_p }A`-`Y^C{\_m }_EB``Y^E{\_n }_{\_p }A`\big)+\\
&-`\Theta_A^{\_m }`\breve t`{}_{\_m }\,B^C`+\alpha`\bbP_B^C{\_m }_E`\breve t`{}_{\_m }\,A^E`
-`\Theta_A^{\_m }`\breve Q`{}_{\_m }\,B^C`+\alpha`\bbP_B^C{\_m }_E`\breve Q`{}_{\_m }\,A^E` \,.
\end{formula}

We have already stressed that for double and exceptional field theories the C-constraint is redundant as long as $`F_MN^P`$ does not contain the trombone component.
Because of the constraint \eqref{X-constraint} this is equivalent to asking that the $\bbR^+_0$ component vanishes.
We can actually make a stronger statement, namely that
for these theories the embedding tensor requirement \eqref{C-like constr for X} is redundant as long as $`X_AB^C`$ does not gauge $\bbR^+_0$.
Indeed, in \eqref{explicit F} we can see that if this is the case then $\breve X`{}_AC^C`=0$, which in turn implies $`F_MP^P`=0$, keeping in mind the section constraint and the fact that $Q_m$ is H algebra valued.

The only remaining step of our proof is to show that the Bianchi identity \eqref{BI} is always satisfied.
To do so we first rewrite it in an equivalent form.
Taking \eqref{BI} and contracting with $`E_A^M`$ we notice that one of the derivative terms can be replaced by the expression \eqref{Lie der of F} after setting $\Lambda^M=`E_A^M`$:
\begin{formula}\label{uhk}
`E_A^m`\partial_m`F_NP^Q` 
= \bbL_{E_A}`F_NP^Q`-2`\cE_[M|^m`\partial_m `E_A^T` `F_T|N]^P` - `Y^PS_RN``\cE_S^m`\partial_m`E_A^T``F_TM^R`
\end{formula}
Notice that this identity holds by virtue of the C-constraint.
Performing this substitution we arrive at an equivalent expression for the Bianchi identity:
\begin{formula}\label{BI equivalent}
\widetilde\bbL_{E_A}`F_MN^P` =
\bbT[`\Theta_M^m`\partial_m(`E_A^T``F_TN^P`)]
\end{formula}
where $\bbT$ is a shorthand notation for the torsion projection defined in \eqref{torsion from weitz}, so that $`T_AB^C` = \bbT[`W_AB^C`]$.
In \eqref{BI equivalent} it is understood to act on the indices $MNP$, leaving $A$ as a spectator.

We now use the definition of $`F_MN^P`$ in \eqref{F from T and X} and the property \eqref{SS condition}\footnote{We stress again that \eqref{SS condition} is satisfied \emph{by definition} of $`F_MN^P`$. What we are proving is that such $`F_MN^P`$ satisfies all consistency conditions.} to write the left hand side as
\begin{formula}\label{BI proof lhs}
\widetilde\bbL_{E_A}`F_MN^P` =
`E_M^B``E_N^C``E_D^P`\big(
-\delta_{X_A}`T_BC^D`-`E_A^m`\partial_m`T_BC^D`
\big)\,,
\end{formula}
where $\delta_{X_A}$ is the duality algebra variation under the generator $X_A$.
We may now bring the factors $`E_M^B``E_N^C``E_D^P`$ to the right hand side of \eqref{BI equivalent} and write it as 
\begin{formula}\label{BI proof rhs}
`E_B^M``E_C^N``E_D^P`\bbT[`\Theta_M^m`\partial_m(`E_A^T``F_TN^P`)] =
\bbT\big[\,[W_B,\,E\circ F_A]`{}_C^D`+`E_B^m`\partial_m(E\circ `F_AC^D`)\big]\,.
\end{formula}
The constraint \eqref{X-constraint} together with \eqref{weitz of coset first version} imply that $E\circ `F_AB^F``W_FC^D`=0$, so that the first term can be rewritten as $-\bbT[\delta_{F_A}`W_BC^D`] = -\delta_{F_A}`T_BC^D`$.
The second term reduces to $`E_B^m`\partial_m`T_AC^D`$ and adding back the left hand side of \eqref{BI equivalent} we arrive at
\begin{formula}\label{rvfd}
\delta_{X_A}`T_BC^D`+`E_A^m`\partial_m`T_BC^D`=\delta_{(E\circ F)_A}`T_BC^D`-\bbT[`E_B^m`\partial_m`T_AC^D`]\,.
\end{formula}
Noticing that $`X_AB^C`=`T_AB^C`+E\circ `F_AB^C`$ and expanding $\bbT$, this expression reduces to the property \eqref{BI for torsion} of the torsion $`T_AB^C`$.
This concludes our proof that \eqref{BI equivalent} and hence \eqref{BI} are satisfied.

\subsection{Patching and global extension}\label{global}

We now investigate how the local construction of the previous sections extends globally.
First, we note that if we change our choice of coset representative, $L(y)_A{}^B\to h(y)_A{}^CL(y)_C{}^B$ for some $h(y)\in\Hg$, the reference Vielbein and the frame transform as
\begin{formula}\label{h gauge trf of e and E}
`\oe_m^{\_m }` &\to `\oe_m^{\_n }` \mathring h`{}-1_{\_n }^{\_m }`\,,\\
`E_A^M` &\to `L-1_A^B``h-1_B^C`\mathring h`{}_C^D``\oe_D^M` 
= `E_A^N``q_N^M`\,,\qquad `q_N^M`\in(\cG_0\times\bbR^+_0)\ltimes\cP_0\,,
\end{formula}
where $\mathring h`{}_A^B`$ is the projection of $h(y)$ to \GL(d), so that $`h-1_B^C`\mathring h`{}_C^D` \in (\cG_0\times\bbR^+_0)\ltimes\cP_0$.
The transformation $`q_M^N`$ is then obtained by conjugation with the Vielbein.

Take now two coordinate patches $U_{\mathsf a},\ U_{\mathsf b}$ with coordinates labelled as $y_{\mathsf a}^m,\,y_{\mathsf b}^m$ respectively.
On their intersection $U_{\mathsf a} \cap U_{\mathsf b}$ the coset representatives are related by
\begin{formula}\label{patching of L}
L_{\mathsf b}(y_{\mathsf b}){}_A{}^B = h_{\mathsf{ab}}(y_{\mathsf b}){}_A{}^C L_{\mathsf a}\left(y_{\mathsf a}(y_{\mathsf b})\right){}_C{}^B\,.
\end{formula}
Notice that we are not assuming to have a globally defined coset representative, which would require the possibility to globally remove the H-transformations $h_{\mathsf{ab}}$ from the patching above. 
Leaving the arguments as understood, we thus arrive at the associated patching of the local frame
\begin{formula}\label{patching of E}
`E_{\mathsf{b}}\,A^M` &= (L_{\mathsf a}^{-1})`{}_A^B` h_{\mathsf{ab}}^{-1}`{}_B^C`\mathring h_{\mathsf{ab}}\,`{}_C^D`\oe_{\mathsf a}`{}_D^N``J-1_{\mathsf{ab}}\,N^M`
= `E_{\mathsf a}\,A^P``J-1_{\mathsf{ab}}\,P^N` `q-1_{\mathsf{ab}}\,N^M`\,,
\end{formula}
where $`J_{\mathsf{ab}}\,N^M`$ is the inverse Jacobian of the change of coordinates between the patches, embedded in the duality group as usual.
We thus see that $`E_A^M`$ is patched together by diffeomorphisms and transition functions $`q_{\mathsf{ab}}\,M^N`$ valued in $(\cG_0\times\bbR^+_0)\ltimes\cP_0$.
Cocycle conditions can be traced back (using \eqref{h gauge trf of e and E} in reverse) to the cocycle conditions for the \H-valued transition functions $`h_{\mathsf{ab}}\,A^B`$ and are automatically satisfied.

The fact that the \GL(d) part of the transition function is just the standard change of basis induced by coordinate transformations is essential to be able to consistently patch together the frame.
However, the $q_{\sf ab}$ transformations can be problematic for two reasons and need to be analysed in detail.
We will find that both the following problems can be overcome.
First, a globally defined generalised frame should be patched together by diffeomorphisms and $p$-form gauge transformations.
Instead, $q_{\sf ab}$ appears to take values also in $\cG_0\times\bbR^+_0$. 
Second, the $\cP_0$ part of $q_{\sf ab}$ may not be exact. 
Namely, it may correspond to shifts of the $p$-form potentials by parameters that are not the differential of a $(p-1)$-form.
These issues are reflected in the patching properties of $`F_MN^P`$, which will transform by conjugation with the same transition functions as $`E_A^M`$, plus an inhomogeneous term that is essentially the torsion projection of $`\Theta_M^m`\partial_m q_{\sf ab} q^{-1}_{\sf ab}$.
This is a consequence of the inhomogeneous transformation of $Q_m$ under H.

A conservative solution is to restrict ourselves to embedding tensors $`X_AB^C`$ which do not gauge any $\cG_0\times\bbR^+_0$ generators, so that neither $q_{\mathsf{ab}}$ nor $`F_MN^P`$ contain components in the global symmetries of the higher dimensional theory.
The problem of non-exactness of $q_{\mathsf{ab}}$, if it arises, is solved passing to an \emph{untwisted frame} by extracting a $\cP_0$-valued component from $L$ and absorbing it in $`F_MN^P`$.
This is described more in detail in appendix~\ref{appLdeco}. 
The final result is that we define
\begin{formula}\label{Etild}
\tilde E_A{}^M &\equiv E_A{}^N \tilde C_N{}^M\,,\\
\tilde F_{MN}{}^P &\equiv \tilde C^{-1}\circ F_{MN}{}^P + \bbT[`\Theta_M^m`\tilde C^{-1}{}_N{}^Q\partial_m\tilde C_Q{}^P ]=
\tilde E\circ(X-T[\tilde E])`{}_MN^P`\,,
\end{formula}
where $\tilde C_M{}^N$ is patched by diffeomoprisms and by $q_{\sf ab}$.
These objects automatically satisfy
\begin{formula}\label{untw SS}
\bbL_{\tilde E_A} \tilde E_B{}^M -\tilde E_A{}^P\tilde E_B{}^Q \tilde F_{PQ}{}^M = -`X_AB^C`\tilde E_C{}^M
\end{formula}
and are patched together by $J_{\sf ab}$ exclusively.
This guarantees that they are globally defined sections of appropriate untwisted generalised bundles.
In other words, $\tilde E_A$ is a collection of global vectors and $p$-forms (and possibly $p$-form densities) encoding the background internal metric, warp factor, and scalar fields, while $\tilde F$ encodes background fluxes and massive deformations.
One may then extract the $p$-form fluxes from $\tilde F`{}_MN^P`$ identifying a locally defined $\hat C_M{}^N \in \cP_0$ that encodes the associated $p$-form potentials and that is patched together by $p$-form gauge transformations exclusively,\footnote{Namely, by some $\hat q_{\sf ab}\in\cP_0$ whose associated (local) Weitzenb\"ock connection has vanishing torsion.} so that
\begin{formula}\label{extract Chat}
\hat E_A{}^M \equiv \tilde E_A{}^N \hat C{}_N{}^M\,,\qquad \hat F`{}0_MN^P` \equiv \hat C^{-1}\circ \tilde F`{}_MN^P` + \bbT[`\Theta_M^m`\hat C^{-1}{}_N{}^Q\partial_m\tilde C_Q{}^P ]\,,
\end{formula}
and $\hat E_A$, $\hat F^0$ define a global solution of \eqref{SS condition twisted}.
Now $\hat E_A$ is a global frame for the (twisted) generalised tangent bundle and $\hat F^0$ only encodes massive deformations of the uplift theory, if any are necessary.
We also anticipate that $\hat F^0$ can inform us on whether our uplift can be extended by extra flat directions, as we will discuss in the next section.

Making a few reasonable assumptions on the properties of the supergravities and generalised geometries under consideration, we can extend the discussion of the paragraph above to a wider class of uplifts.
First, we will assume that only a single copy of the (standard) tangent bundle is embedded into the generalised tangent bundle.
In other words, given a generalised vector $V^M$ only its components $V^M`\Theta_M^m`$ transform as a vector under \GL(d).
This also means that an object $B_M$ transforms as a one-form only if $B_M = `\Theta_M^m`B_m$.
Second, we point out that in most supergravity theories only the scalar currents and their dual $D+d-2$ forms transform in the adjoint of the global symmetry group $\cG_0\times\bbR^+_0$.
This implies that the representation content of $`F_MN^P`$ can only allow for terms valued in $\cG_0\times\bbR^+_0$ that are either \GL(d) singlets, corresponding to an embedding tensor in the higher dimensional theory, or \GL(d) one-forms corresponding to fluxes induced by global symmetry twists.
Both these assumptions apply for instance to maximal supergravities and to the associated exceptional generalised geometries.
Finally, we are going to focus on uplifts to supergravities that are not themselves gauged.
This allows us to discuss global definiteness without the need to worry about gauge-group valued transition functions.
In principle, this is a requirement that we could lift.

To avoid the uplift theory to be itself a gauged supergravity we just need to exclude components of $`X_AB^C`$ that are valued in the Lie algebra of $\cG_0\times\bbR^+_0$ and are \GL(d) singlets.
As a result any $\cG_0\times\bbR^+_0$ valued components of $`F_MN^P`$ must be the torsion projection of a one-form $`\Theta_M^m``B_mN^P`$.
A close look at \eqref{explicit F} shows that any such contributions coming from $\breve t_{\underline m}$ cancel out with $\breve X`{}_AB^C`$.\footnote{The term $\alpha`\bbP_P^Q{\_m }_R`\breve t`{}_{\_m }\,M^R`$ is valued in $\cP_0$.}
Moreover, the H-valued part of $\breve X`{}_AB^C`$ cannot contribute because it would not correspond to a \GL(d) one-form.
This implies that actually, under our current assumptions
\begin{formula}\label{H in GLdP0}
\H\subset\GL(d)\ltimes\cP_0\,.
\end{formula}
Thus we conclude that $`F_MN^P`$ does not contain any components valued in the Lie algebra of $\cG_0\times\bbR^+_0$ and that $q_{\sf ab}\in\cP_0$, which brings us back to the procedure described above to construct a globally defined frame.
Of course, if some of our assumptions are not satisfied for more exotic generalised geometries we can always impose \eqref{H in GLdP0} directly.

We should stress that the conditions for global definiteness discussed above are only required if we want our frame to be global in the sense of standard generalised geometries, where transition functions for the generalised tangent bundle are not valued in $\cG_0\times\bbR^+_0$.
We may however be willing to relax this requirement.
For instance, it appears perfectly acceptable to have $\cG_0\times\bbR^+_0$ valued transition functions if the higher dimensional theory is itself gauged. 
This is taken into account by our construction in presence of \GL(d) singlet components gauging $\cG_0\times\bbR^+_0$ in $X_A$, which become directly part of $\hat F^0$ and determine the higher dimensional gauging. 
Even if we were to allow for arbitrary twists taking values in the global symmetry group of the higher dimensional theory, the resulting geometries would appear less pathological than general U-folds, in which the patching is performed by duality transformations that are not global symmetries.

\subsection{Central charges and extended internal space}\label{extension}

Our final objective is to lift the assumption made below \eqref{vectoralgebra}, where we restricted the vector components of the frame to be of the form $`\Theta_A^a`K_a$.
In doing so we will be able to capture the most general instance of generalised Leibniz parallelisable spaces and the associated generalised frames.

It is instructive to first consider an intermediate step, namely an easy extension of the situation described so far, in  which  the generalised flux constraints \eqref{X-constraint}, \eqref{C-constraint} and \eqref{BI} are satisfied by $\hat F^0$ on a space larger than $\H\backslash\G$.
To check if this is the case we introduce an extended section matrix 
\begin{formula}\label{larger sec matrix}
`\cE_M^{\hat m}`\equiv (`\Theta_M^m`,\,`\cE_M^{m_0}`)\,,\quad m_0=d+1,\ldots,d+n\,.
\end{formula}
and require that it solves the section constraint \eqref{SCa} as well as all the generalised flux constraints \eqref{X-constraint}, \eqref{C-constraint} and \eqref{BI} with $\hat F^0_{MN}{}^P$ in place of $`F_MN^P`$.
Notice that this is now a requirement on $`\cE_M^{m_0}`$.
If we find such a non-vanishing extension of the section matrix, then the internal space is extended to
\begin{formula}\label{extended int space}
\H\backslash\G\times\bbT^n\,,
\end{formula}
possibly with some warping of the $\bbT^n$ factor over the coset space.
This does not require a modification of the frame and fluxes, which therefore do not depend on the torus (angle) coordinates.
It is also worth noticing that there can be more than one such extension. 
For instance, uplifts of maximal gauged supergravities on coset spaces of low dimension might be extendable to both type IIB or eleven dimensional supergravity depending on a choice of $`\cE_M^{m_0}`$.

Let us now complete our analysis and consider the most general situation that can arise from \eqref{vectoralgebra}, in which other vectors $K_A$, not proportional to $`\Theta_A^a`K_a$, are allowed to be non-vanishing.
In this case \eqref{vectoralgebra} tells us that the $K_A$ define a centrally extended version of the G algebra.
The situation described in the paragraph above is a special case, where all the non-vanishing vectors independent from $K_a$ sit in the right kernel of $h_{ab}{}^{a_0}$, thus determining an extension of G by direct product with $\U(1)$ factors.
We will denote the centrally extended gauge group $\G_{\rm ext}$ (direct product or not), which is not a subgroup of $\cG\times\bbR^+$, and refer to the central extension as $\mathrm Z$, so that $\G=\G_{\rm ext}/\mathrm Z$.
As in section~\ref{recipe}, we notice that $\G_{\rm ext}$ acts transitively on the internal space, which is thus a coset space $\H_{\rm ext}\backslash\G_{\rm ext}$.
Because the central charges commute with everything else, we can \emph{locally} reduce this coset space to
\begin{formula}\label{reduce extended coset space}
\H_{\rm ext}\backslash\G_{\rm ext} \simeq \H\backslash\G \times\ \bbT^n\,,
\end{formula}
where $\bbT^n$ denotes the coset space directions associated with central charges and $\H = \H_{\rm ext}/(\H_{\rm ext}\cap \mathrm Z)$.
Globally, the torus fibration over $\H\backslash\G$ can be non-trivial.

We may now define an extended embedding tensor $\hat\Theta_A{}^{\hat a}$, where $\hat a=(a,\,a_0)$ runs along a basis for the right kernel of $`X_(AB)^C`$, so that the following requirements are satisfied:
\begin{formula}\label{frcd}
`X_(AB)^C`\hat\Theta_C{}^{\hat a} &= 0\,,\qquad
\hat\Theta_A{}^{a} = \Theta_A{}^{a}\,,\qquad
\hat\Theta_A{}^{a_0} \perp \Theta_A{}^{a} \ \ \forall a_0\neq a\,,
\end{formula}
and so that we can write the most general $K_A$ satisfying \eqref{vectoralgebra}  as
\begin{formula}\label{qgtf}
K_A = \hat\Theta_A{}^{\hat a} K_{\hat a}\,,
\end{formula}
with $K_{\hat a}$ being the Killing vectors on the coset space. 
In other words, $\hat\Theta_A{}^{\hat a}$ defines the embedding of the adjoint of $\G_{\rm ext}$ into the $\*R _{\rm v}$ indices and is $\G_{\rm ext}$ invariant.
We can now repeat the same argument used in section \ref{recipe} to arrive at \eqref{frame Ansatz} and \eqref{F from T and X}, with \eqref{reduce extended coset space} instead of $\H\backslash\G$ and $\hat\Theta$ instead of $\Theta$.
In particular, we need to require that the projection $\hat\Theta_A{}^{\hat{\_m }}$ of the extended embedding tensor onto a set of coset generators $t_{\hat{\_m }}$ satisfies the section constraint
\begin{formula}\label{ext theta SC}
`Y^AB_CD`\hat\Theta_A{}^{\hat{\_m }}\hat\Theta_B{}^{\hat{\_n }} = 0\,.
\end{formula}
The extra linear constraint \eqref{C-like constr for X} must also be amended by substituting $\Theta_B{}^{\_m }\to\hat\Theta_B{}^{\hat{\_m }} $.
An important observation is that Z is trivially represented in the $\*R _{\rm v}$ representation, so that $`L_A^B`$ will still be a coset representative of $\H\backslash\G$, in particular $`L_A^B`\in\G\subset\cG\times\bbR^+$.
Instead, the reference Vielbein $\oe_{\hat m}{}^{\hat{\_m }}$ might be non-trivial along the torus directions, depending on whether the Cartan--Maurer equations $\mathrm{d}\Omega+\Omega\wedge\Omega=0$, projected onto the generators of $\bbT^n$ translations, imply that the associated one-form is locally exact or not.
If it is, we may use the expressions developed for the $\H\backslash\G$ truncation and treat the  torus extension as at the beginning of this section.
If it is not, we must take into account that now $\oe\in\GL(d+n)$ and that $(\cG_0\times\bbR^+_0)\ltimes \cP_0$ are the global symmetries and $p$-form transformations of the theory living on the $d+n$ dimensional internal space.
The resulting $`E_A^M`$ and $`F_MN^P`$ will differ from those we would have obtained by uplifting on $\H\backslash\G$ exclusively.\footnote{An uplift on $\H\backslash\G$ is always guaranteed, as we can enlarge $\H_{\rm ext}$ to include the whole Z, and the section constraint as well as \eqref{C-like constr for X} will be automatically satisfied by the resulting $`\Theta_A^{\_m }`$ if they were for $\hat\Theta`{}_A^{\hat{\_m }}`$.}
In any case, under an $\H_{\rm ext}$ transformation both $`L_A^B`$ and $`\oe_A^M`$ will only transform with $\H$, which is now a subgroup of $(\GL(d+n)\times\cG_0\times\bbR^+_0)\ltimes\cP_0$.

At this point, the proofs in section~\ref{proof} must be repeated.
All the steps turn out to be exactly the same if we simply make the substitutions
\begin{formula}\label{subsextproof}
d\to d+n\,,\qquad
`\Theta_A^a`\to\hat\Theta`{}_A^{\hat a}`\,,\qquad
`f_ab^c`\to \delta^a_{\hat a}\delta^b_{\hat b}(`f_ab^c`\delta_c^{\hat c}+`h_ab^{c_0}`\delta_{c_0}^{\hat c})\,.
\end{formula}
As regards the global patching discussed in section~\ref{global}, the transition functions $q_{\sf ab}$ are still induced by H transformations because Z acts trivially on both the coset representative and the reference Vielbein.
Except for the dimensionality of the internal space which is enlarged to $d+n$, and the substitution $\Theta\to\hat\Theta$ there are no differences in the analysis.
The untwisted and the twisted generalised frame and fluxes are constructed using the same procedure.

This completes our constructive proof that the most general generalised Leibiniz parallelisable space is of the form \eqref{reduce extended coset space} with $\G_{\rm ext}$ the central extension of the gauge group G determined from $`X_AB^C`$ itself.\footnote{The algebra of $\G_{\rm ext}$ could be regarded as the maximal Lie subalgebra of the Leibniz algebra with structure constants $`X_AB^C`$.}

\section{Examples}
\label{examples}

\subsection{Group manifold reductions}

Our procedure includes group manifold reductions as a special case. 
Suppose $\Gg=\G'\ltimes \Hg$ with $\G'\subset\GL(d)$.
In this case the reference Vielbein $`\oe_m^{\_m }`$ is the right invariant Vielbein on $\G'$ and \eqref{frame Ansatz} reduces to the (inverse of the) left invariant Vielbein on $\G'$, embedded into the duality group.
Thus, \eqref{frame Ansatz} becomes the standard Scherk--Schwarz reduction Ansatz on group manifolds written in the language of ExFT/EGG.
It is guaranteed to generate upon truncation a gauged supergravity with constant embedding tensor $`X0_AB^C`$.
If $`X0_AB^C`\neq`X_AB^C`$, the difference is entirely generated by background fluxes, massive deformations or gaugings of the higher dimensional theory reduced on $\G'$. 
This last statement is non-trivial and is a consequence of the general proof of consistency of section~\ref{proof}.

\subsection{Consistent Pauli reductions}

The consistent Pauli reductions on group manifolds $\G'$ discussed in \cite{Baguet:2015iou} are consistent truncations of double field theory that map the complete set of isometries $\G=\G'_{\rm left}\times\G'_{\rm right}$ of a group manifold $\G'$ into the gauge group of the reduced theory.
In this setting the duality group is $\cG=\O(d,d)$ with invariant metric $\eta_{AB}$, $d$ being the dimension of the group manifold and $A,\,B,...$ being \O(d,d) vector indices.
The duality group admits subgroups $\SO(p,q)\times\SO(p,q)$ with $p+q=d$ so that $A=(i,{\_i })$, each set of indices being the vector irrep of one \SO(p,q) factor.
The structure tensor is $`Y^AB_CD`=\eta^{AB}\eta_{CD}$ and the general solutions of the section constraint span a subspace of the vector representation that is null with respect to $\eta_{AB}$.
The embedding tensor is a set of structure constants $`f_AB^C`$ with only nonvanishing components $`f_ij^k`=`f_{\_ij }^{\_k }`$ corresponding to two copies of the $\G'$ structure constants.
The Pauli reduction on $\G'$ is based on the equivalent coset space
\begin{formula}\label{pauli coset}
\G'\simeq \frac{\G'_{\rm left}\times\G'_{\rm right}}{\G'_{\rm diag}}\,.
\end{formula}
We can indeed take $\Hg=\G'_{\rm diag}$ which is automatically embedded in the \GL(d) subgroup of \SO(d,d). 
The coset generators are then taken to be the anti-diagonal combination of left and right generators, which transform in the adjoint of \Hg.
Projecting $`f_AB^C`$ onto the anti-diagonal combinations we obtain a set of null vectors $`f_A^{\_m }`$ that satisfy the section constraint.
No $\cG_0\times\bbR^+_0$ components are present and hence we obtain a global Leibniz parallelisation.

\subsection{$\omega$-deformed \SO(p,q) gaugings and a no-go result}

Let us now focus on certain classes of gaugings of four-dimensional maximal supergravity described in \cite{DallAgata:2011aa,DallAgata:2012bb,DallAgata:2014tph}.
They include as special cases the original \SO(8) gauging of de Wit and Nicolai \cite{deWit:1981sst,deWit:1982bul} and the non-compact \SO(p,q) and contracted \CSO(p,q,r) gaugings of \cite{Hull:1984vg,Hull:1984qz} (see \cite{deWit:2007kvg} for a treatment based on the embedding tensor and \cite{Trigiante:2016mnt} for a review of the whole subject).

The discussion of the latter gaugings also applies to the similar families that exist in other dimensions.
The four dimensional case is however richer because of the presence of symplectic deformations \cite{DallAgata:2012bb,DallAgata:2014tph} which allow for inequivalent choices of the embedding tensor sharing the same gauge group, but giving rise to different physics.\footnote{It would certainly be interesting to perform a similar analysis for the symplectic deformations of the half-maximal gauged supergravities discussed in \cite{Inverso:2015viq}, making use of SL(2)-DFT \cite{Ciceri:2016hup}.}

We start from the gauged maximal supergravities with $\Gg=\SO(8)$ \cite{deWit:1981sst,deWit:1982bul,DallAgata:2012bb}.
The only possible coset space with dimension low enough to satisfy the section constraint is $\SO(7)\backslash\SO(8)$.
There are three inequivalent subgroups $\SO(7)_{v,s,c}$ depending on which of the three irreps $\mathbf8_v,\,\mathbf8_s,\,\mathbf8_c$ decomposes into $\*7 +\*1 $.
We also know that for the section constraint \eqref{sec constr Theta} to be satisfied the embedding of \SO(7) into $\Eseven\times\bbR^+$ must go through the chain
\begin{formula}\label{SO7embchain}
\SO(7)\subset\GL(7)\subset\Eseven\times\bbR^+\,.
\end{formula}
This rules out one of the three choices, say $\SO(7)_s$, because it goes through an embedding in \SU(7) rather than \GL(7).\footnote{We are picking conventions in which the fermions transform in the $\*8 _s $ and $\*56 _s$ representations of \SO(8).}
The other two choices satisfy the embedding chain.
In fact, $\SO(7)_v$ and $\SO(7)_c$ are mapped into each other by an \Eseven transformation that normalises \SO(8) \cite{DallAgata:2011aa,DallAgata:2014tph}.

All inequivalent embedding tensors for \SO(8) are parameterised by an angle $\omega\in[0,\,\pi/8]$ \cite{DallAgata:2012bb}.
Other values of $\omega$ are equivalent to those in the specified range. 
Working in the standard \SL(8) symplectic frame (see e.g. \cite{deWit:2007kvg}), the original \SO(8) gauging of de Wit and Nicolai is obtained for $\omega=0\mod\pi/4$.
It is described by a purely electric embedding tensor $`\Theta_\Lambda^\alpha`$ where $\Lambda=1,\ldots,28$  only runs along the `electric' half of the $\*56 $ irrep of \Eseven.
All other gaugings are obtained by turning on a magnetic component proportional to the electric one, the relative coefficients being specified by $\omega$.
We can write the electric embedding tensor replacing the adjoint index $\alpha$ with the adjoint of \SO(8) and using double index notation (each couple corresponds to one of the two indices of $\Theta$):
\begin{formula}\label{SO8ambtensdoubleindex}
`\Theta_\Lambda^\alpha` \sim `\delta_[{\_A }^[{\_C }``\delta_{\_B }]^{\_D }]`\,,\qquad {\_A },\,{\_B },\,{\_C },\,{\_D } = 1,\ldots,8\,.
\end{formula}
The \SO(8) generators are also written as $t_{[{\_AB }]}$.
Let us say that ${\_A }$ is an index in the $\*8 _v$ irrep.
The coset space $\SO(7)_v\backslash\SO(8)=S^7$ is generated by $t_{[{\_A8 }]}$.
The expression corresponding to $`\Theta_A^{\_m }`$ becomes
\begin{formula}\label{seccondSO8solv}
`\Theta_A^{\_m }` \sim `\delta_[{\_A }^[{\_m }``\delta_{\_B }]^{\_8 }]`\,,\qquad{\_m }=1,\ldots,7\,,
\end{formula}
which indeed solves the section constraint and reproduces eleven-dimensional supergravity~\cite{Hohm:2013uia}.
Because $\Hg\subset\GL(7)$ exclusively, there are no issues with the global extension of the generalised frame.
Actually, the frame $`E_A^M`$ matches the untwisted frame $\tilde E`{}_A^M`$ and $`F_MN^P`=\tilde F`{}_MN^P`$ encodes the Freund--Rubin flux.

We can also pick the coset space $\SO(7)_c\backslash\SO(8)=S^7$, but the electric embedding tensor above will not solve the section constraint once we project onto the new coset generators.
Because $\SO(7)_c\simeq\SO(7)_v$ by \Eseven conjugation, we can conjugate the embedding tensor by the same transformation exchanging the two \SO(7) groups \cite{DallAgata:2014tph}.
The resulting embedding tensor corresponds to $\omega=\pi/4$ and does solve the section constraint when projected onto the generators of $\SO(7)_c\backslash\SO(8)=S^7$.
More importantly, these are the only combinations of isotropy group and $\omega$ deformation that satisfy the uplift conditions.

This analysis of the \SO(8) case may appear redundant, because we knew from the start that the $\omega=\pi/4$ theory is equivalent to the standard one, and hence equally liftable to 11d supergravity.
It has also been shown that the other inequivalent \SO(8) gaugings ($\omega\in(0,\,\pi/8]$) do not admit a geometric uplift to 11d supergravity \cite{deWit:2013ija,Lee:2015xga}.
The information we have just collected is however crucial for the non-compact cousins of the $\SO(8)_\omega$ gaugings, where the no-go theorem of \cite{Lee:2015xga} does not apply.

Let us first look at $\SO(4,4)_\omega$, which are especially interesting because of their family of de Sitter extremal points that can satisfy arbitrary slow-roll conditions by tuning the value of $\omega$ \cite{DallAgata:2012plb}.
These gaugings have a structure very similar to \SO(8) and actually all the discussion above applies directly.
In particular, the $\SO(4,4)_{\pi/4}$ gauging is equivalent to $\SO(4,4)_0$ \cite{DallAgata:2014tph} and there are two subgroups $\SO(4,3)_{v,c}$ analogous to $\SO(7)_{v,c}$ and related by an \Eseven transformation \cite{DallAgata:2011aa,DallAgata:2014tph}.
We do not find an uplift for any other value of $\omega$.

For other signatures the story diverges in some small but relevant ways.
The value $\omega=\pi/4$ is inequivalent to $\omega=0$ for \SO(7,1), $\SO(6,2)\simeq\SOs(8)$ and \SO(5,3).
All these theories have vacua of some kind when setting $\omega=\pi/4$ \cite{DallAgata:2011aa}, as well as other solutions, some even supersymmetric, for varying values of $\omega$ \cite{Borghese:2012qm,Borghese:2013dja,Gallerati:2014xra}.
In particular, $\SO(6,2)_{\pi/4}$ is the starting point to construct a large family of theories exhibiting Minkowski vacua with varying amounts of residual supersymmetry \cite{DallAgata:2011aa,Catino:2013ppa}.
All these gauge groups have subgroups analogous to $\SO(7)_v$ and $\SO(4,3)_v$ which allow to uplift the $\omega=0$ variants as done in \cite{Hohm:2014qga,Baron:2014bya}.
To uplift the $\omega=\pi/4$ theories, though, we would need a subgroup analogous to $\SO(7)_c \subset\SO(8)$, e.g. an $\SO(6,1)$ or $\SO(5,2)$ subgroup of \SO(6,2) such that the spinorial $\*8 _c$ branches to $\*7 +\*1 $.
Unfortunately, these real forms do not admit such subgroups and the outer (triality) automorphism mapping vector and spinorial irreps is broken by the choice of real section.
Other values of $\omega$ are excluded by the section constraint as usual.

We now combine these results with two observations. First, the \SO(p,q) algebras are simple and thus do not admit central extension and have faithful adjoint representation. Second, their embedding in \Eseven is such that no element of the irrep  $\*R _{\rm v}=\*56 $ is invariant.
The consequence of these observations is that any frame $\hat E_A$ must have vector components $\hat E`{}_A^M``\cE_M^m`=`\Theta_A^a``K_a^m`$ with $`K_a^m`$ satisfying the \SO(p,q) algebra.
Referring back to our discussion in section~\ref{recipe}, this means that the procedure we have just followed to look for uplifts is exhaustive.
The results of the previous paragraph therefore imply the following no-go result
\begin{center}\it\raggedright
The only $SO(8)_\omega$ and $SO(p,q)_\omega$ gaugings admitting a (locally or globally) geometric uplift are the undeformed ones ($\omega=0$ or equivalent).
{\widowpenalties 1 10000 \par}
\end{center}
The first part of our proof is analogous to \cite{Lee:2015xga}, but then we do not need to rely on the existence of an invariant generalised metric or of a maximally (super)symmetric vacuum solution, which were restricting the the no-go result stated there to the compact case $\SO(8)_\omega$.

\subsection{\CSO(p,q,r) gaugings revisited}

Let us now move to the gaugings of $\ISO(7)=\CSO(7,0,1)=\SO(7)\ltimes\bbR^7$.\footnote{It is entirely trivial to change the signature to \ISO(p,q) and we will not discuss this further.}
There are two such gaugings \cite{DallAgata:2014tph}.
The first one is entirely electric, has no vacua and uplifts to massless type IIA supergravity on $S^6$ \cite{Guarino:2015jca,Guarino:2015qaa,Guarino:2015vca}.
The second one has an embedding tensor equal to the first, plus an extra magnetic contribution by a term which reproduces exactly the Romans mass deformation $\hat F^0$ of the generalised Lie derivative for massive IIA supergravity \cite{Ciceri:2016dmd}.
Put in these terms, it will not come as a surprise that such gauging lifts to massive IIA on a six-sphere \cite{Guarino:2015jca,Guarino:2015qaa,Guarino:2015vca}.\footnote{Chronologically, however, the uplift \cite{Guarino:2015jca,Guarino:2015qaa,Guarino:2015vca} came before an XFT/EGG for massive type IIA was formulated \cite{Ciceri:2016dmd,Cassani:2016ncu} and used complementary techniques analogous to those of the $S^7$ consistent truncation \cite{deWit:1986oxb}.}
In the approach of this paper, the six-sphere uplifts just mentioned are obtained from the coset space
\begin{formula}\label{redcosISO7sphere}
\frac{\ISO(7)}{\ISO(6)\times\bbR}\ =\ \frac{\SO(7)}{\SO(6)}\ =\ S^6\,.
\end{formula}
For the electric \ISO(7) gauging we could actually pick $\frac{\ISO(7)}{\ISO(6)}$ as seven-dimensional internal space and the extra flat direction would correspond to a standard Kaluza--Klein compactification from eleven-dimensional supergravity to massless type IIA.
The final expression for $\hat E$ would not differ from \eqref{redcosISO7sphere} and the extra flat direction is recovered from \eqref{redcosISO7sphere} as an $S^1$ extension allowed by the vanishing of $\hat F^0$ (again, only for the electric gauging).
Similar ambiguities will apply to the choice of internal space for the other \CSO(p,q,r) gaugings discussed below and we choose to always display the most economical coset space.

Because the Romans mass deformation only affects the gauge connection of the $\bbR^7$ generators modded out in \eqref{redcosISO7sphere}, it does not affect the construction of $`E_A^M`$ in any way, but rather passes trough the entire uplift procedure and becomes the $\hat F^0$ deformation of the EGG Lie derivative as anticipated.
This is entirely consistent with the analysis of \cite{Guarino:2015vca}.

For the electric \ISO(7) gauging there is another choice of  coset space.
This time we pick $\Hg=\SO(7)\subset\GL(7)$ and keep the seven translations, so that the internal space is just $\bbR^7$ and the uplift is to eleven-dimensional supergravity.
All consistency conditions are satisfied, including global definiteness on $\bbR^7$.
It is also straightforward to find out that $`F_MN^P`=0$, so that there are no background $p$-form fluxes or other deformations.
The generators of $\bbR^7$ are embedded into a subalgebra of \eseven which is the \emph{transpose} of $\mathfrak p_0$.
In the language of \cite{Aldazabal:2010ef}, this means that the background under consideration is a realisation of \ISO(7) in terms of a `locally geometric flux' on a flat internal space.
The nomenclature refers to the fact that upon compactification on $\bbT^7$ the supergravity fields will jump along cycles by U-duality transformations.
The fact that \ISO(7) can be recovered both as a sphere reduction and as a locally geometric flux on a torus exemplifies once more the known fact that the interpretation of embedding tensor components as geometric or non-geometric is devoid of meaning until we fix our choice of uplift Ansatz.
If we do not compactify to $\bbT^7$ the metric and $C_3$ field blow up at infinity.
This is an immediate consequence of the exponential dependence of $`E_A^M`$ on the Cartesian coordinates of $\bbR^7$.

We can repeat the analysis of \ISO(7) for the other \CSO(p,q,r) groups.
For each one of them we find the uplift manifolds of \cite{Hohm:2014qga} and an uplift on flat internal space with some `locally geometric flux' analogous to the one found for \ISO(7).\footnote{This becomes the standard NSNS $Q$-flux of ten-dimensional supergravities for \CSO(2,0,6) and \CSO(1,1,6).}
Moreover, the \CSO(3,0,5) gauging can not only be uplifted on $S^2\times\bbT^5$ as done in \cite{Hohm:2014qga}, and to flat space as just specified.
It can also be uplifted to an $S^3$ group manifold in the standard Scherk--Schwarz fashion.
In total, counting only the most economical coset spaces as discussed for ISO(7), this gauging admits three inequivalent uplift manifolds\footnote{For the distinction between $S^3$ and $S^3/\bbZ_2$ see the comments section.}
\begin{formula}\label{CSO305 manifolds}
\frac{\CSO(3,0,5)}{\CSO(2,0,5)\times \bbR^5}= S^2\,;\quad
\frac{\CSO(3,0,5)}{\CSO(3,0,4)}=\bbR^3\,;\quad
\frac{\CSO(3,0,5)}{\bbR^{15}}\simeq S^3\,.
\end{formula}
The non-compact version also works similarly.

For all coset spaces described above $`\Theta_A^{\_m }`$ is a submatrix of \eqref{seccondSO8solv} and therefore solves the section constraint.
The only exceptions are the $S^3$ group manifold reduction in \eqref{CSO305 manifolds} and its non-compact version, where $`\Theta_A^{\_m }`\sim \delta_{\_AB }^{\_np }\epsilon^{\_npm }$.
This also satisfies the section constraint.\footnote{There are actually some sign flips in $`\Theta_A^{\_m }`$ when dealing with the non-compact versions, which we have been ignoring in our exposition.}
In all the cases with $r\ge2$, the generalised flux constraints allow to extend the internal space with extra flat directions to reach uplifts to both type IIB and eleven dimensional supergravity.

All coset spaces in this section can be reduced to have $\H\subset\GL(d)$.
This implies that $`E_A^M`$ is the untwisted frame of the reduction and that a global extension is ensured.
Excluding the cases with generalised $Q$-flux and the group manifold, the other uplifts will include a $d$-form flux embedded into $`F_MN^P`$.
This is entirely analogous to the expressions already known in the literature and we do not discuss it further.

The `dyonic' CSO gaugings of \cite{DallAgata:2011aa} can be also uplifted following our procedure.
These are superpositions of two \CSO(p,q,r) groups taking the form
\begin{formula}\label{dyCSO}
(\SO(p,q)\times\SO(p',q'))\ltimes N
\end{formula}
with $N$ generated by a nilpotent algebra.
The story is entirely similar to the discussion above, except that the coset space will decompose into two pieces, corresponding to the electric and magnetic parts of the gauging.
This is consistent with the uplift expressions developed in \cite{Inverso:2016eet}.
Beyond the uplifts described there, it is straightforward to deduce from the section constraint that semisimple \SO(3) and \SO(2,1) factors can also be uplifted as group manifolds, and that uplifts of either CSO copy based on locally geometric fluxes are only allowed when $p+q+p'+q'<8$.\footnote{Notice that we can also make choices for $\H$ such as $\H=(\SO(p,q-1)\times\SO(p',q'))\ltimes N$ so that we only uplift one of the two CSO copies while the other becomes a gauging for the higher-dimensional theory.}

\section{Comments}\label{outro}

We have identified a general procedure to uplift gauged supergravities in terms of generalised Leibniz parallelisations for the associated ExFT.
Consistency requires that we find a subgroup $\H$ of the gauge group \G such that the projection $`\Theta_A^{\_m }`$ of the embedding tensor on a set of $\H\backslash\G$ coset generators $t_{\_m }$ satisfies the section constraint \eqref{sec constr Theta}, and if necessary the extra linear constraint \eqref{C-like constr for X}.
If central extensions are present, the same constraints apply to the extended embedding tensor $\hat\Theta`{}_A^{\hat{\_m }}`$ described in section~\ref{extension}.

There is in principle an alternative way to check whether a certain gauged supergravity admits an uplift based on our construction.
This is worth mentioning as it exemplifies the difficulty in generating a generalised Leibniz parallelisation with embedding tensor $`X_AB^C`$ if we choose a solution of the section constraint $`\cE_M^m`$ which is not tailored to the embedding tensor.
Let us first choose the target higher-dimensional theory and an associated solution of the section constraint $`\cE_M^m`$.
Then we can define a projector $\Pi_M{}^N$ onto the $d$ dimensional vector space defined by $`\cE_M^m`$.
A certain embedding tensor admits an uplift following our procedure if the projected gauge generators $x_A\equiv(\bbone-\Pi)`{}_A^B`X_B$ form a Lie subalgebra of the gauge algebra.
Namely, $x_A$ must satisfy the quadratic constraint \eqref{QC} and define H.
The linear constraint \eqref{C-like constr for X} must also be imposed.
The disadvantage of this approach is that $`\Pi_A^B`$ is not unique nor covariant under $\cG\times\bbR^+$ and the procedure above must be repeated for the whole $\cG\times\bbR^+$ orbit of $`X_AB^C`$ and for each choice of $`\Pi_A^B`$ (although the orthogonal projector is likely to be the correct guess, as it is in all our examples). 
Letting the embedding tensor induce by itself the correct choice of solution of the section constraint as done in the main text is an important technical simplification which is made possible by the ExFT formalism.

It is natural to ask under what conditions the generalised parallelisations described here are well-defined once we quotient the internal space by some group of discrete isometries.
Because in \eqref{frame Ansatz} we use the coset representative written in the $\*R _{\rm v}$ representation of $\cG$, it is clear that the natural global versions of $\G$ and $\H$ to be considered are the ones faithfully represented in $\*R _{\rm v}$.
We have implicitly used this argument to describe the central extensions in section~\ref{extension} as $\bbT^n$ rather than $\bbR^n$.
Another simple example is the uplift on $S^7=\SO(7)\backslash\SO(8)$.
In this case our procedure really identifies $\SO(7)\backslash\PSO(8)=\bbR\bbP^7=S^7/\bbZ_2$ as the internal space, because the $\bbZ_2$ center of \SO(8) is trivially represented on $\*R _{\rm v} \overset{\SO(8)}{\to} \*28 +\*28 $.
Thus the generalised frame is automatically well-defined on $\bbR\bbP^7$, which is consistent with the counting of supersymmetries in \cite{Duff:1997qz} and with the supersymmetry enhancement of the ABJM model at level $k=2$ \cite{Aharony:2008ug}.
Of course the extension to the double cover $S^7$ is straightforward.
Notice that the same discussion applies to the $\mathrm{GL}^+(d+1)$ generalised parallelisation of any odd-dimensional sphere and its $\bbZ_2$ quotient $\bbR\bbP^d$.
These observations are already useful to identify some allowed global forms of the internal spaces.
Whether extra quotients can be allowed would be an interesting question to investigate.
It would require the presence of further (discrete) isometries beyond those in G.
While we do not rule out entirely that some extra quotients exist, the expectation is that actions by transformations non-trivially represented in $\*R _{\rm v}$ will require $\cG\times\bbR^+$ valued transition functions and thus define U-fold like geometries.

We have left out of our discussion the $D=3$ $\mathrm E_{8(8)}$ ExFT \cite{Hohm:2014fxa,Baguet:2016jph}.
In three dimensions dual graviton contributions enter prominently in the algebra of generalised diffeomorphisms, which does not close \cite{Berman:2012vc} unless extra covariantly constrained gauge parameters are introduced \cite{Hohm:2014fxa}.%
\footnote{These extra parameters can be gauged-fixed introducing a preferred connection in the generalised Lie derivative \cite{Rosabal:2014rga,Cederwall:2015ica}, although this connection does not match with the one appearing naturally in the supersymmetrisation \cite{Baguet:2016jph}.}
Generalised Scherk--Schwarz reductions for $\mathrm E_{8(8)}$ ExFT have not been discussed in the literature yet.
Recent progress has been made in \cite{Hohm:2017wtr} by constructing an half-maximal \O(d+1,d+1) ExFT in three dimensions conceptually analogous to the four-dimensional \SL(2)-DFT recently developed in \cite{Ciceri:2016hup}.
It appears likely that an approach similar to the one followed in this paper will work for the construction of generalised Scherk--Schwarz reductions of three-dimensional ExFTs.
A natural first step in this direction is the construction of the most general flux deformations of the generalised Lie derivative in $D=3$.

We have briefly explained how our recipe reproduces the known uplifts of many gauged supergravities, provides a few alternative uplifts for some, and excludes a geometric origin for others.
The natural next step is to exploit this formalism to generate new uplifts of gauged supergravities and use them to construct new interesting solutions of string- and M-theory.
It would be particularly interesting to investigate whether there exist any generalised Leibniz parallelisable spaces belonging to  the larger class presented in section~\ref{extension}, with non-trivial fibration over $\H\backslash\G$.
We hope to come back to these questions in the near future.

\bigskip

{\ }\\

\noindent\textbf{\underline{Note added after publication}}\ 
An important typo in the position of indices $I$ and $H$ in the uplift constraint \eqref{C-like constr for X} has been corrected, which propagated from the same expression also repeated in equations \eqref{C-like for torsion} and \eqref{wesdtcfgh}. 
An erratum for the journal version of this article has been published. 
I thank the authors of \cite{Bugden:2021wxg} for discussions that led me to notice this problem.

\section*{\href{https://www.youtube.com/watch?v=v8ahNggrc_Y}{A}cknowledgements}
I thank Gianguido Dall'Agata for discussions and early collaboration on related topics, and John Huerta for discussions.
I thank Franz Ciceri and Adolfo Guarino for useful comments on a draft version of this paper.
This work is partially supported by FCT/Portugal through a CAMGSD post-doc fellowship.

\appendix
\section{Coset representative decomposition}\label{appLdeco}

Any element of $\cG\times\bbR^+$ can be decomposed in terms of a compact transformation $U\in \cH$ and elements of $(\GL(d)\times\cG_0\times\bbR^+_0)\ltimes\cP_0$.
This is so because the supergravity degrees of freedom parameterising $\cG/\cH$ all descend from the internal metric, scalar fields of the higher dimensional theory (e.g. dilaton or axio-dilaton) and $p$-forms.
Because the coset representative of $\H\backslash\G$ is embedded in $\cG\times\bbR^+$, we can apply the same decomposition:
\begin{formula}\label{L deco}
[L^{-1}]`{}_M^N` = [U\,G\,S\,P]`{}_M^N`\,,
\end{formula}
where $G\in\GL(d)$, $S\in\cG_0\times\bbR^+_0$ and $P\in\cP_0$.
The components $G$ and $S$ can be modified by \O(d) and $\cH_0$ transformations that can be reabsorbed into $U$, but we will not make direct use of this fact.
$P$ is unambiguously identified.

A similar decomposition applies to H transformations, where a $\cH$ element is not required
\begin{formula}\label{h deco}
[h^{-1}]_M{}^N = [\mathring h^{-1}\, s\, p]_M{}^N
\end{formula}
where $\mathring h \in\GL(d)$, $s \in \cG_0\times\bbR^+_0$ and $p\in\cP_0$.
On an overlap between two patches, $L$ transforms by such an H transformation.
Substituting \eqref{h deco} into \eqref{L deco} we can deduce the transition functions for each factor
\begin{formula}\label{deco trf properties}
U_{\sf b}  &=   U_{\sf a}\,(G_{\sf a} \mathring h^{-1}_{\sf ab} G_{\sf b}^{-1})( S_{\sf a} s_{\sf ab} S^{-1}_{\sf b} ) \,,    \\
G_{\sf b}  &=  (G_{\sf a} \mathring h^{-1}_{\sf ab} G_{\sf b}^{-1})^{-1} G_{\sf a} \mathring h_{\sf ab}^{-1}      \,,\\
S_{\sf b}  &=  ( S_{\sf a} s_{\sf ab} S^{-1}_{\sf b} )^{-1} S_{\sf a} s_{\sf ab}      \,, \\
P_{\sf b}  &=  P_{\sf a} ( P^{-1}_{\sf a} \mathring h_{\sf ab} s^{-1}_{\sf ab} P_{\sf a}s_{\sf ab} \mathring h^{-1}_{\sf ab} p_{\sf ab} )       \,.
\end{formula}

In section~\ref{global} we have described conditions under which $s_{\sf ab}$ is trivial.
In such situation we may define
\begin{formula}\label{Ltild}
\tilde L^{-1} \equiv L^{-1} P^{-1}\,,
\end{formula}
which is patched together by $\mathring h^{-1}_{\sf ab}$ exclusively.
The transition functions $q_{\sf ab}$ become the conjugation by $\oe$ of the transition function for $P$.
Equivalently, defining $\tilde C\equiv \oe P^{-1} \oe^{-1}$ we arrive at the untwisted frame 
\begin{formula}\label{untw frame appendix}
\tilde E \equiv \tilde L^{-1} \oe^{-1} = E \tilde C \,.
\end{formula}
The local twist $\tilde C$ is patched with the same $q_{\sf ab}$ transistion functions that appear in $E$ so that they cancel out in the product and $\tilde E$ is patched together by internal diffeomorphisms exclusively, consistently with the patching deduced from \eqref{Ltilde}.
If $s_{\sf ab}$ is non-trivial, the above definitions are still valid but $\tilde E$ and $\tilde F$ will be patched together with extra $\cG_0\times\bbR^+_0$ transformations.

%
%

\small

\providecommand{\href}[2]{#2}\begingroup\raggedright\endgroup

\end{document}